\documentclass[aip,amsmath,amssymb,reprint]{revtex4-1}

\usepackage{graphicx}
\usepackage{dcolumn}
\usepackage{bm}
\usepackage{float}
\usepackage{color}
\usepackage{xcolor}
\usepackage{color}
\usepackage{soul}
\usepackage{titlesec}
\usepackage[nottoc]{tocbibind}
\usepackage[normalem]{ulem}

\usepackage[utf8]{inputenc}
\usepackage[T1]{fontenc}
\usepackage{mathptmx}

\draft 

\begin{document}

\title{Understanding developing turbulence by a study of the nonlinear energy transfer in the Navier-Stokes equation} 

\author{Preben Buchhave}
\email[]{buchhavepreben@gmail.com}
\affiliation{Intarsia Optics, S{\o}nderskovvej 3, 3460 Birker{\o}d, Denmark}

\author{Clara M. Velte}
\email[]{cmve@dtu.dk}
\homepage[Personal web page: ]{https://www.staff.dtu.dk/cmve}
\homepage[Projects web page: ]{https://www.trl.mek.dtu.dk/}
\affiliation{Department of Civil and Mechanical Engineering, Technical University of Denmark, Niels Koppels All\'{e}, Building 403, 2800 Kongens Lyngby, Denmark}

\date{\today}

\begin{abstract}
    In the present work, we investigate a numerical one-dimensional solver to the Navier-Stokes equation that retains all terms, including both pressure and dissipation. Solutions to simple examples that illustrate the actions of the nonlinear term are presented and discussed. The calculations take the full 4D flow as its starting point and continuously projects the forces acting on the fluid at a fixed Eulerian point in a stationary coordinate system onto the direction of the instantaneous velocity. Pressure is included through modeling. Adhering to the requirement that time must in general be considered an independent variable, the time development of the time records and power spectra of the velocity fluctuations are studied. It is found that the actions of the nonlinear term in the Navier-Stokes equation manifests itself by generating sharp pulses in the time traces, where the sharpness is bounded by the finite viscosity. In the spectral domain, the sharp gradients in the pulses generate energy contributions at high frequencies that yields a $-2$ slope across the inertial range. The $-2$ (or $-6/3$) slope is explained through a simple example and the classically expected $-5/3$ slope in the inertial range can be recovered from the pressure fluctuations from the full flow field that can be considered a noise contribution at the point considered. We also observe that the spectrum can in principle keep spreading to higher frequencies or wavenumbers without upper bound, as the viscosity is approaching the zero limit.\\
    \keywords{Turbulence, Non-equilibrium turbulence, Triad interactions, Non-local interactions, Navier-Stokes Equation, Non-linear Differential Equations, Burger's equation, Fractal grids, Dissipation Anomaly, Permanence of large eddies}
\end{abstract}

\maketitle 



\section{\label{sec:introduction}Introduction}
The Navier-Stokes equation is a nonlinear partial differential equation of motion describing the momentum of a fluid flow at a point in space as a function of time. This equation is generally accepted to describe all aspects of the momentum balance of Newtonian constant density fluid flows, both laminar and turbulent. If supplemented with equation of state, it also describes compressible flows and non-Newtonian flows. However, due to the nonlinearity of the equation and the wide range of scales interacting through the nonlinear term, the Navier-Stokes equation is notoriously difficult to solve and requires considerable computer power to evaluate numerically, especially for turbulent flows at large Reynolds numbers~\cite{1,2}.

To throw some light on the properties of the Navier-Stokes equation, one-dimensional equations with some of the same properties as the Navier-Stokes equation have been studied, e.g. Burgers' equation~\cite{3,3b}, with and without the diffusion term, and Euler's equation. Euler's equation is interesting for the study of the influence of the nonlinear convection term, but it is unable to describe the time development of a realistic, random velocity field. In a comprehensive review~\cite{4}, the common properties of Burgers' equation and the Navier-Stokes equation and their differences were discussed. It was pointed out that Burgers' equation also lacks the sensitivity to boundary and initial conditions that are inherent in the Navier-Stokes equation~\cite{5}.

In the current work, we investigate a one-dimensional numerical solution to the Navier-Stokes equation representing the full momentum balance for the instantaneous fluid convection velocity~\cite{6}. The calculations retain all terms in the Navier-Stokes equation, including both pressure and dissipation. By projecting all forces acting on the fluid control volume (CV) onto the direction of the instantaneous three-dimensional instantaneous velocity vector, we retain the full four-dimensional dynamics of the problem and are not limited to a single spatial dimension and time. The pressure is included through modeling, and the fluctuating pressure gradient at the measuring point adds turbulent energy to the velocity field.

Our method allows us to follow an arbitrary input velocity record as it develops in time by following it through a large number of infinitesimal passes (iterations) through a 1D projection of the force terms in the Navier-Stokes equation onto the instantaneous flow direction. This method, if applied with sufficiently small steps, will build up the nonlinear effects, in principle without approximation. The 1D projection means that we are measuring the spatial velocity structures as they are convected through the CV. The method does not allow us to obtain certain features of the large-scale 3D spatial flow field, but it does allow computation of, e.g., the kinetic energy, the velocity power spectrum in both the spatial and the temporal domain as well as calculations of the $2^{nd}$ and $3^{rd}$ order structure functions in the flow direction (c.f. Buchhave and Velte\cite{6} and Zhu \textit{et al.}\cite{6a}). As a result, we can study the so-called triad interactions between Fourier components of the spectral representation of the flow field and follow their interactions through both space and time. The method has been compared and validated with measurements, c.f. the work of Dotti \textit{et al.}\cite{7} and other results presented in section~\ref{sec:singleFmodeInjection}.

The purpose of this paper is primarily to learn about the peculiar effects of the nonlinearity of the Navier-Stokes equation by presenting some calculations of the flow development with different representative input velocity records. The triad interactions and their dynamics may be said to form the basic machinery of fluid flow evolution, and a detailed understanding of their function is important for both theory and modelling of turbulence. We thus nicknamed the program the ``Navier-Stokes Machine''.

\begin{figure}[t]
    \centering
    \includegraphics[width=\columnwidth]{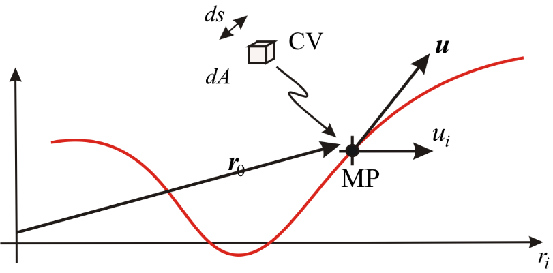}
    \caption{Control volume CV, instantaneous velocity u and a component $u_i$ at the CV, a spatial point at position $r$ and at time, $t$. The red line indicates the path of a fluid element in time (streak line). Color figure available online.}
    \label{fig:1}
\end{figure}


The Navier-Stokes equation for normalized density ($\rho=1$) in the physical space coordinate system,
\begin{equation}\label{eqn:1}
\frac{\partial \mathbf{u}(\mathbf{r},t)}{\partial t} = - \mathbf{u}(\mathbf{r},t) \cdot \nabla \mathbf{u}(\mathbf{r},t) - \nabla p(\mathbf{r},t)+\nu \nabla^2 \mathbf{u}(\mathbf{r},t),
\end{equation}
expresses the time rate of change of momentum of a fluid element, a control volume, at position $\mathbf{r}= (r_1, r_2, r_3)$ with velocity $\mathbf{u}(\mathbf{r},t) = \left ( u_1(\mathbf{r},t), u_2(\mathbf{r},t), u_3(\mathbf{r},t)\right )$ at time $t$. The first term on the right-hand side of the normalized equation represents convective acceleration, the second fluctuating pressure force and the third viscous action due to momentum diffusion.

For simplicity, we consider constant density flows, which means we can set the divergence of the velocity equal to zero:
\begin{equation}\nonumber 
\nabla \cdot \mathbf{u}(\mathbf{r},t) =0
\end{equation}

Pressure plays a special role in the Navier-Stokes equation. All other terms act locally on an infinitesimal fluid parcel as illustrated by the control volume in Figure~\ref{fig:1}, whereas pressure is a 3D effect that must be evaluated in the whole fluid volume, including boundary effects. These effects are not included in Burgers' equation~\cite{5}. Knowledge of the pressure is needed to describe the large-scale behavior of the flow, such as vortices and coherent structures, which are associated with spatial pressure gradients. The fluctuating pressure gradient, originating from velocity fluctuations elsewhere in the flow, adds energy to the small and intermediate scales, the velocity power spectrum and the spatial structure functions, whose time development are governed by the local nonlinear term in the Navier-Stokes equation, as will be demonstrated in the current work.

In addition to pressure differences, many other processes can add energy to a flow: Shear layers such as e.g. boundary layers and free shear layers, pump action, objects in the flow path etc. The immediate effect is to create an imbalance between the spatial velocity structures (``eddies'') containing the turbulent energy and the smaller high frequency velocity structures. If the flow is then left unforced, the imbalance will eventually develop into a flow with equilibrium between the eddies, where energy is transferred predominantly from large to small eddies until eddies become so small that the energy is converted to heat due to the viscosity of the fluid (the so-called Richardson-Kolmogorov cascade). Global equilibrium across all scales can also be achieved for continuously forced turbulence, so that a balance between forcing and dissipation is eventually achieved, and the spectrum retains its shape and energy. In the present work, we have chosen to consider only initial forcing of a homogeneous velocity field and study the development of the resulting decaying flow.

The exact process from non-equilibrium to equilibrium flow is one of the least understood features of turbulence today, but also one of the most important to understand, both from a theoretical point of view and from a practical one, since it is of crucial importance in the development of engineering models.

The goal of the research behind this paper is to explore the action of the Navier-Stokes equation by numerical and laboratory experiments designed to expose the energy flow in developing turbulence. It is designed to map out the path of the energy flow in developing turbulence from the large energy carrying eddies to smaller eddies and eventually to dissipation and to measure the important characteristic time constants for the interactions between turbulent structures. The experiments are designed to clearly show the ``machinery'' of the Navier-Stokes equation acting on turbulent spatial structures. In a companion paper, we describe laboratory experiments where a single large eddy (Fourier component) is introduced in a controlled flow, and the development of the frequency components of the developing velocity power spectrum are measured as a function of time~\cite{7}. Later in the present paper, we compare these experimental results to computations with the Navier-Stokes Machine program using the initial measured record as input to the algorithm. In another paper, we consider the effect of performing a four-dimensional Fourier transform on the flow velocity recognizing the fact that the velocity fluctuations are in reality a stochastic function of four independent variables, three spatial variables and time~\cite{8}.

In the following, we explain the underlying ideas behind the Navier-Stokes Machine computer program. The program can run based solely on variables in physical space (velocity, spatial velocity gradients, pressure gradients etc. within the CV). However, since the structure of the interactions between various spatial velocity scales (``eddies'') is more clearly revealed in Fourier space (or ``$k$-space''), we have developed a version of the program operating on the Fourier components of the velocity. We can then directly see which Fourier components are interacting as the flow evolves. Since the nonlinear term in the Navier-Stokes equation is of the second order, the interactions of Fourier components will always happen between three components, two Fourier components corresponding to two incident k-vectors, say $\mathbf{p}$ and $\mathbf{q}$, and a resulting Fourier component of the spatial frequency $\mathbf{k} = \mathbf{p} + \mathbf{q}$, the so-called triad interactions.

In subsequent sections, we provide some examples of the simulator operating on different samples of input velocity records. The first case treats a deterministic pulse with a broad range of frequencies (a broad range of ``scales'') as input. This example illustrates clearly the effect of the nonlinearity of the Navier-Stokes equation without the disturbing effect of random fluctuations. We then follow up with an input record simulating a stochastic velocity trace modelled with a low frequency van K\'{a}rm\'{a}n spectrum. This example illustrates the transformation from a non-equilibrium incident velocity trace to a fully developed time trace and power spectrum. We then present cases where the input consists of measured velocity traces where a single velocity eddy (Fourier component) is injected by various means into a low and a high turbulence intensity flow, respectively. These examples allow us to clearly show how individual Fourier components develop as a function of time under the influence of the Navier-Stokes equation. Finally, we show a couple of examples illustrating the process of nonlocal interactions. We then discuss what we have learned from the computer experiments and to which extent these results apply to predictions in fully three-dimensional turbulence.

\section{\label{sec:1}Program Description}
The Navier-Stokes equation in its Eulerian form is a differential equation describing the effect of forces acting on a small control volume (CV). It can be described as the change of momentum, $\Delta \mathbf{B}(t)$, or change in velocity, $\Delta \mathbf{u}(t)$, when normalized with a constant density, as a result of external forces, $\sum \mathbf{F}(t)$, acting on the CV during a small time increment, $\Delta t$. Thus, the Navier-Stokes equation in this form is an equation describing not a global field, but rather an equation describing what happens at an Eulerian point in the flow.

When, in the following discussion of the simulation, we speak of ``velocity'', we refer to a digital one-dimensional signal, which we imagine could represent an actual velocity. Referring to Figure~\ref{fig:1}, we notice that, if at any time, $t_0$, we  know the instantaneous velocity vector, $\mathbf{u}(t_0)$, we can simplify the three-dimensional Navier-Stokes equation by projecting the forces onto the direction of $\mathbf{u}$ and onto a plane normal to $\mathbf{u}$. The forces acting in the direction of $\mathbf{u}$ will change the velocity magnitude (and thereby the kinetic energy), while the forces acting in the direction normal to $\mathbf{u}$ will change the direction of $\mathbf{u}$. As the velocity record passes through the CV, we can measure the small-scale velocity structures and compute the temporal and spatial velocity power spectra and the spatial structure functions. In the limit of infinitely small time increments, the contribution to the change in momentum and velocity is linearized and the contributions can be computed individually and added in any order.

\begin{figure}
  \includegraphics[width=\columnwidth]{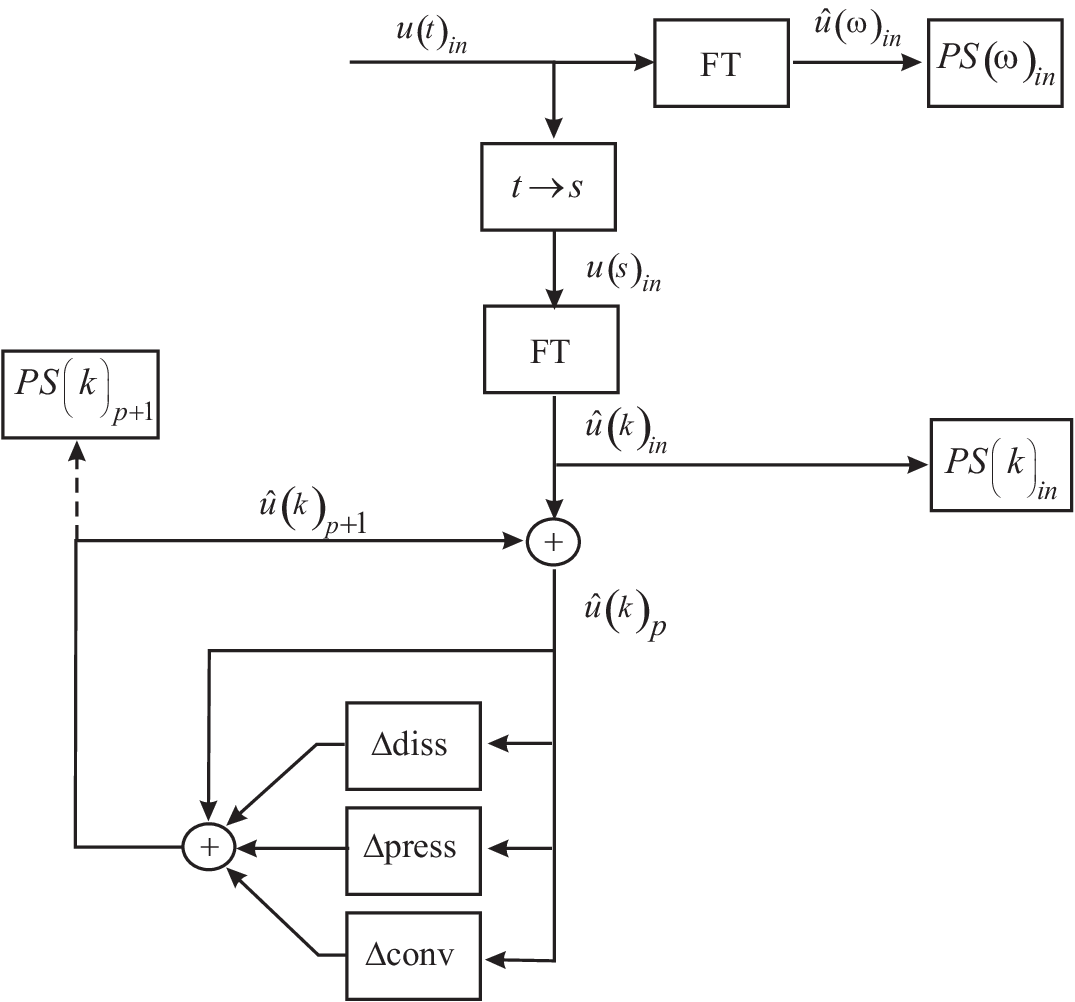}
\caption{Block diagram of the algorithm applied in the current work. }
\label{fig:2}       
\end{figure}

Figure~\ref{fig:2} is a block diagram of the program operating in $k$-space. The input is a one-dimensional velocity time record, $u(t_{n})_{in}$ of length, $T$, consisting of $N$ digital samples enumerated by the index $n$. This record may be created as a digital record in the computer or it could be a digital record obtained from e.g. a laser Doppler anemometer or a hotwire anemometer measurement. Initially, we may compute the input temporal velocity power spectrum, $PS(\omega)_{in}$. As the next step, we convert the velocity time record to a spatial record compensated for the distortions in the sampling process caused by the fluctuating convection velocity~\cite{9}. This record, which we call the convection record, $u(s_n)_{in}$~\cite{6}, can be computed because we know the magnitude of the fluctuating velocity time record, $|u(t_n)_{in}|$:

\begin{equation}\nonumber 
u(s_n)_{in} = \sum_{n'}^{n}|u(t_{n'})_{in}|\Delta t_{n'}
\end{equation}

We then create the spatial Fourier transform, $\hat{u}(k_n)_{in}$, of the spatial input record and, if desired, display the spatial input spectrum, $PS(k_n)_{in}$.

The input velocity record now enters the loop. Each of the three contributions to the changes in the velocity record $\hat{u}(k_n)_{p}$ are iteratively computed and added to create $\hat{u}(k_n)_{p+1}$. The three terms are:
\begin{itemize}
  \item Dissipation: $\Delta diss_p = \nu k_n^2 \hat{u}(k_n)_p  \Delta t$
  \item Pressure: $\Delta press_p = \lambda k_n \hat{u}_e (k_n)_{p}  \Delta t$
  \item Convection: $\Delta conv_p = \sum_{k_1=0}^{N} -i k_1 \hat{u}(k_1)_p  \hat{u}(k_n-k_1)_p\Delta t$
\end{itemize}
The new letters introduced here are $p$, which counts the iterations, the kinematic viscosity of the working fluid, $\nu$, a constant adjusting the strength of the pressure effect, $\lambda$, and the wavenumber, $k_n$, spanning all scales defined by the spectral range. $\hat{u}_e(k)$ in the pressure contribution describes the fluctuating velocity (in wavenumber space) external to the CV. The model for the pressure contribution is described in section~\ref{sec:pressure}.

The process is repeated until a desired view of the changes to the time trace or power spectrum is obtained. The spatial power spectrum and corresponding spatial (or temporal) record can be displayed after a suitable number of passages, $p$, through the computation. To speed up the program, array processing is applied where possible.

The program is a realistic simulation of a process actually taking place in nature. This puts certain restrictions of the possible choices of input signals, physical constants, size of the CV and on the time steps that can be applied in the computation. One can, simply put, say that parameters that could not occur in nature cannot be allowed in the program either; otherwise, the program would simply crash!

\section{\label{sec:2}Examples}
To illustrate the new insights about the nature of the nonlinear spectral transfer, as well as some other interesting observations, we have chosen to present some carefully chosen examples to clearly highlight the desired effects.  

\subsection{Single, short pulse}\label{sec:GaussianPulse}

\begin{figure}[t]
  \includegraphics[width=\linewidth]{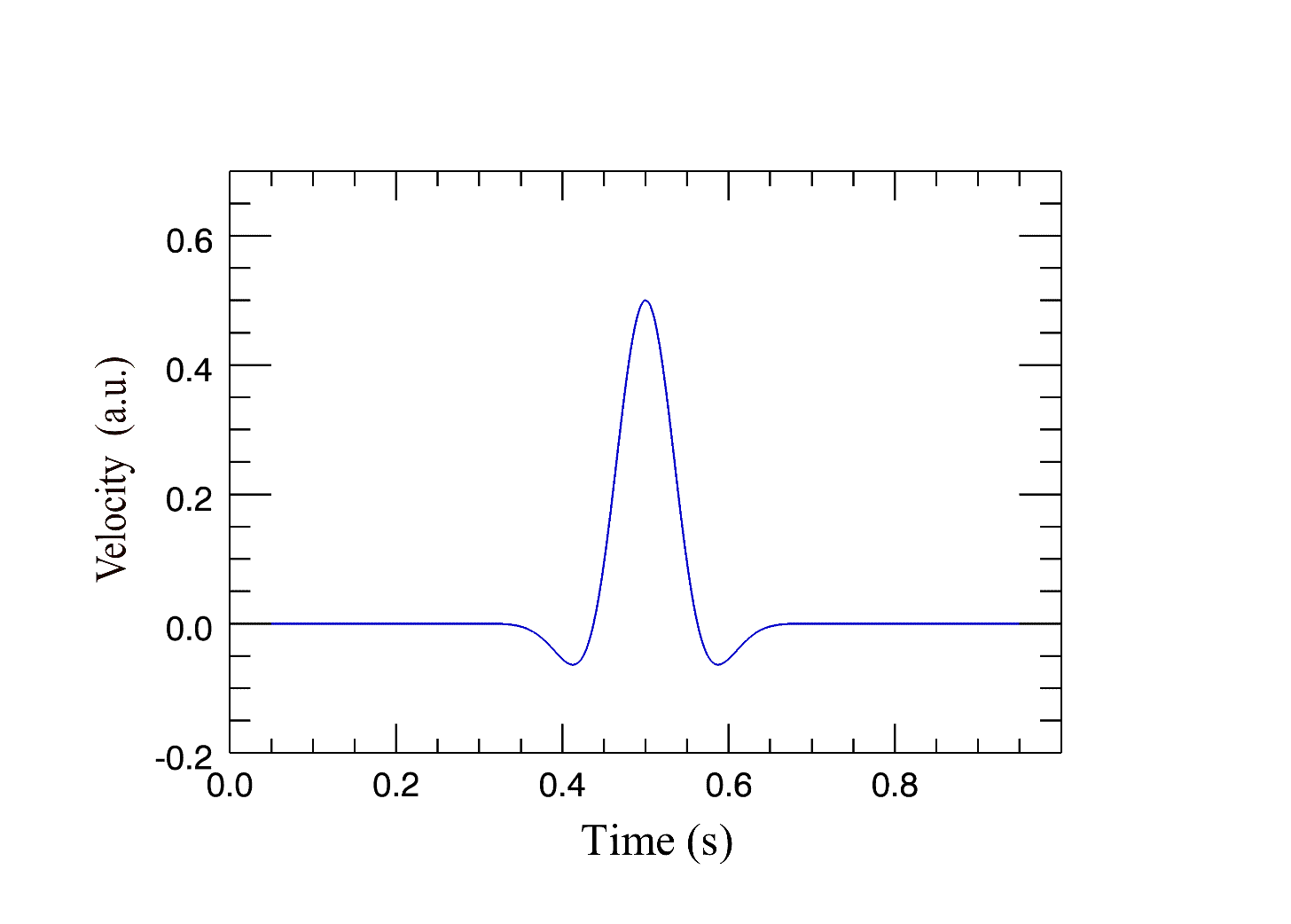}
\caption{Gaussian input time signal. Velocity and other derived quantities in the following is given in arbitrary units (a.u.). Color figure available online. }
\label{fig:3}       
\end{figure}

\begin{figure}[t]
  \includegraphics[width=\linewidth]{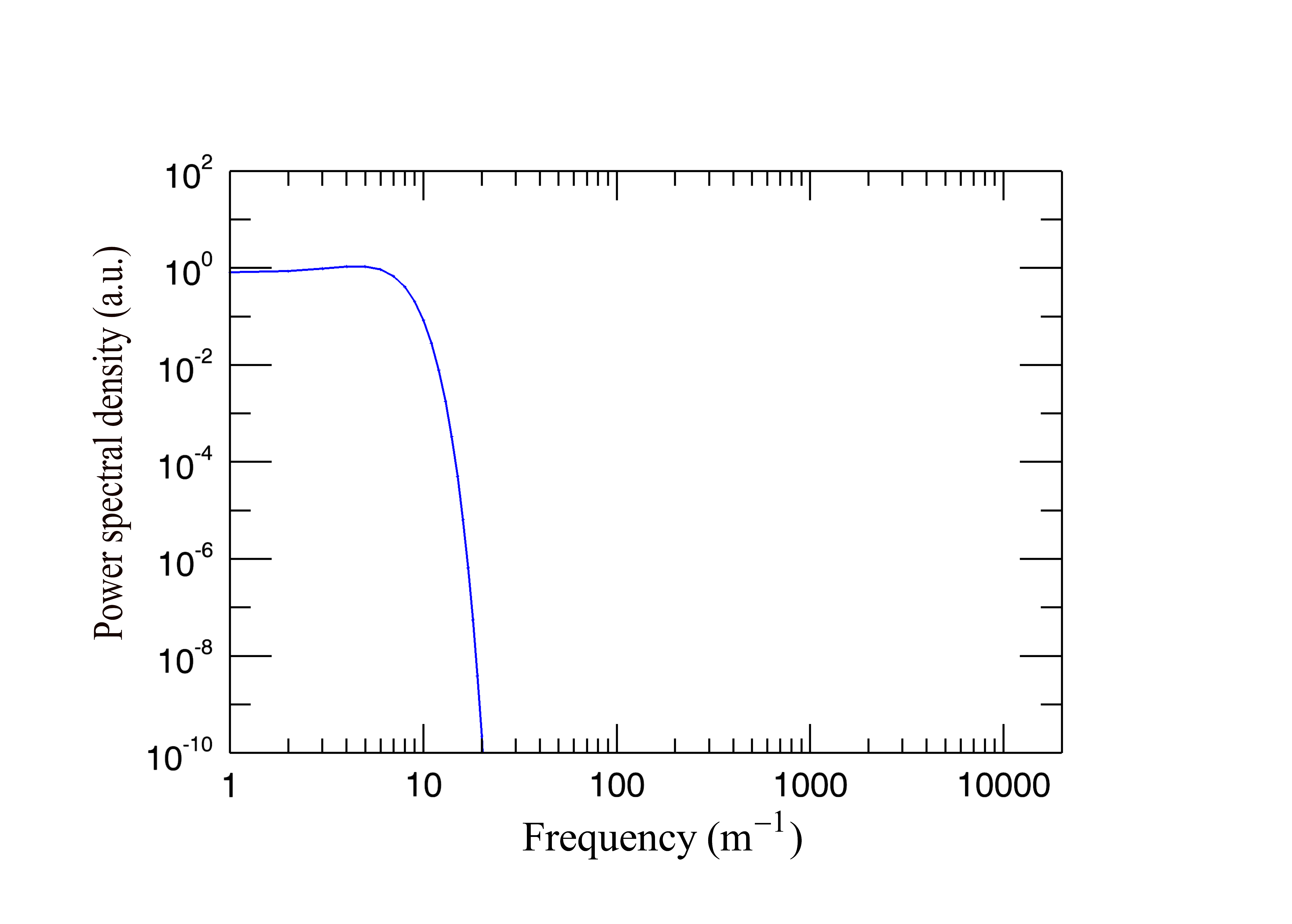}
\caption{Gaussian input spatial frequency spectrum. Color figure available online. }
\label{fig:4}       
\end{figure}

\begin{figure}
  \includegraphics[width=\linewidth]{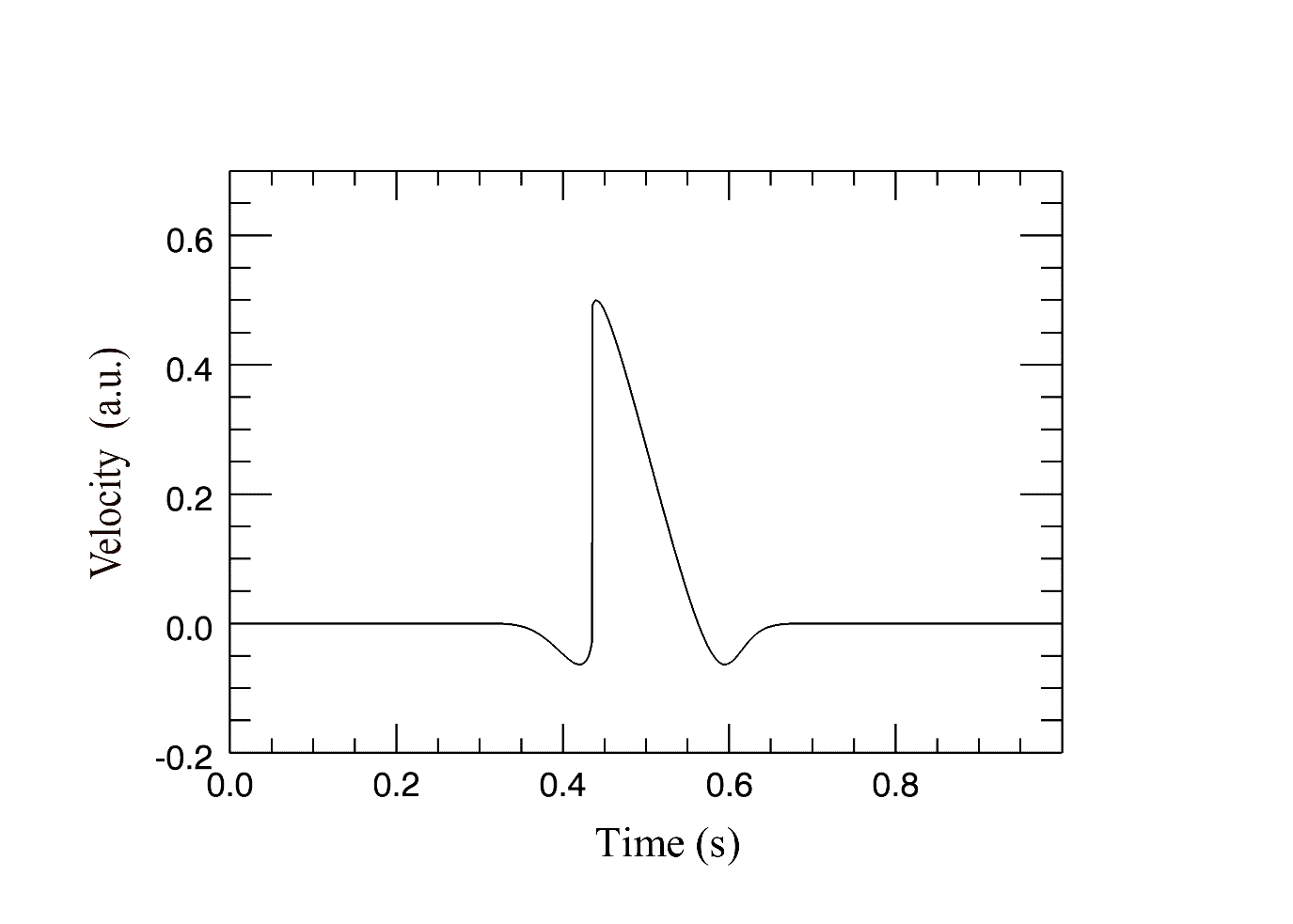}
\caption{Time signal after 12.000 iterations. Video (showing steps of 1.000 iterations) available at https://doi.org/10.11583/DTU.12016866.v1 }
\label{fig:5}       
\end{figure}

\begin{figure}
  \includegraphics[width=\linewidth]{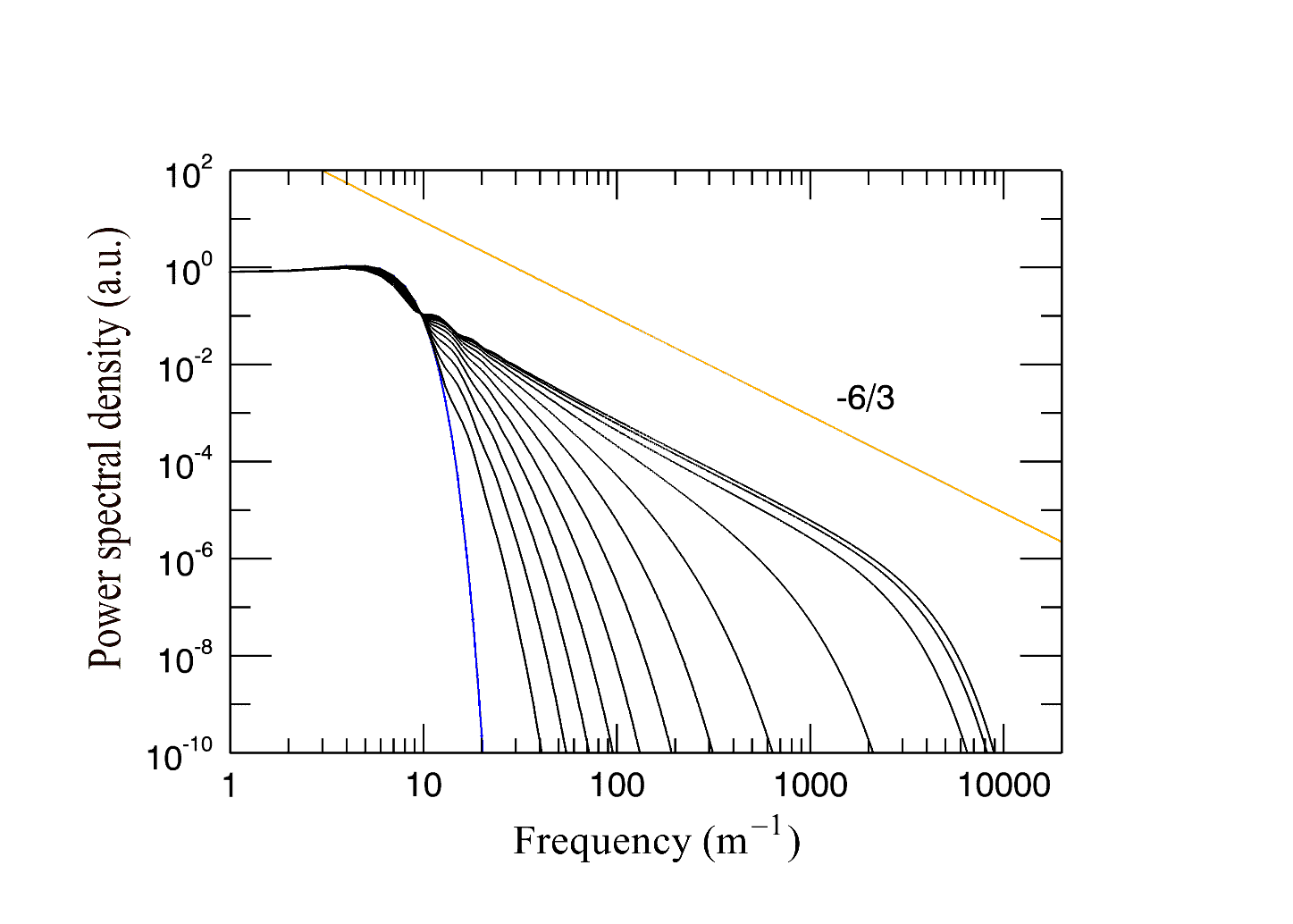}
\caption{Spatial power spectrum after 12.000 iterations (in steps of 1.000 iterations). Color figure available online. Video available at https://doi.org/10.11583/DTU.12016845.v1 }
\label{fig:6}       
\end{figure}

This example shows some of the essential features of the Navier-Stokes equation nonlinearity. We assume a single, 1 second long, time record consisting of a short, Gaussian pulse, (Figure~\ref{fig:3}) which contains a broad spectrum (Figure~\ref{fig:4}). The convection record has been used to obtain the spatial spectrum from the temporal signal. The program includes convection and dissipation, but not pressure. As can be seen in Figure~\ref{fig:5} and Figure~\ref{fig:6} (movies showing steps of 1.000 iterations up to a total of 12.000 iterations), the pulse gets steeper and the power spectrum broadens towards higher spatial frequencies. The power law is cut off at high frequency due to the finite dissipation, which in the time record in Figure~\ref{fig:5} corresponds to the bounded steepness of the pulse (the soft upper edge).

As seen in the figures and also known from studies of Burgers' equation, the shape of the velocity time trace tends towards a `sawtooth' function. The particular form of the nonlinear term, $u\cdot \nabla u$ , can be interpreted as the product between a signal, $u$, and its slope, $\nabla u$. The tendency for the sawtooth function to form can be qualitatively understood from the product of the signal with its slope over repeated actions of the nonlinear term. The sawtooth function can, disregarding the soft upper shock edge due to the finite dissipation, be well approximated by a triangular function $u(s)=\frac{h}{L}(L-s)$, where $2L$ is the pulse length and $h$ its height. The Fourier transform of this function, $\hat{u}(k)=2 \frac{h}{k} \sin (kL) $, will have a $k^{-1}$ dependency and, consequently, the power spectrum $S(k)=\frac{\hat{u}\cdot \hat{u}^{\ast}}{2L}=2 \frac{1}{k^2} \frac{h^2}{L} \sin ^2 (kL) $ will roll off as $k^{-2}$. An analogous analysis can be carried out in the temporal domain. Even the fundamental frequency from the $\mathrm{sin}^2$-function, corresponding to the pulse length, is visible in the spectrum of Figure~\ref{fig:6}. Although the pulses in a measured record are more irregular, the tendency towards triangular pulses will persist, which explains the $-2$ (or $-6/3$) slope of the spectrum.

As will be seen in the third example, on the effect of pressure fluctuations, the classically (or from plain dimensional analysis) expected $-5/3$ slope in turbulence measurements can be attained if the pressure fluctuations are included. But the exact pressure contribution will depend on the flow properties and boundary conditions.

The special form of the nonlinear term is also responsible for the tendency of the kinetic energy to move towards higher frequencies. A simple example illustrates the effects: Assume two co-parallel spatial velocity waves with wavenumbers $k_1$ and $k_2$ propagating along the convection record coordinate $s$, $\cos (k_1 s) + \cos (k_2 s)$. The convection term acting on this function results in the nonlinear product of the velocity after passing through a CV of width $\Delta s$ is then
$$\left ( \cos (k_1 s) + \cos (k_2 s) \right ) \cdot \frac{d}{ds} \left ( \cos (k_1 s) + \cos (k_2 s) \right ).$$
Working out the terms, we get:
$$\frac{1}{2}  [ -k_1 \sin (2 k_1 s) - k_2 \sin (2 k_2 s) $$
$$- (k_1 + k_2) \sin \left ((k_1 + k_2) s \right ) - (k_1 - k_2) \sin \left ((k_1 - k_2) s \right )  ]$$

We see four different frequencies generated from the original $k_1$ and $k_2$: The double frequency of each components and the sum and difference frequencies with the sum-frequency having the greatest coefficient, especially for $k_1 \cong k_2$. Energy will therefore on average tend towards higher values since local interactions have a higher interaction efficiency (as described in~\cite{8}).

\begin{figure}
  \includegraphics[width=\linewidth]{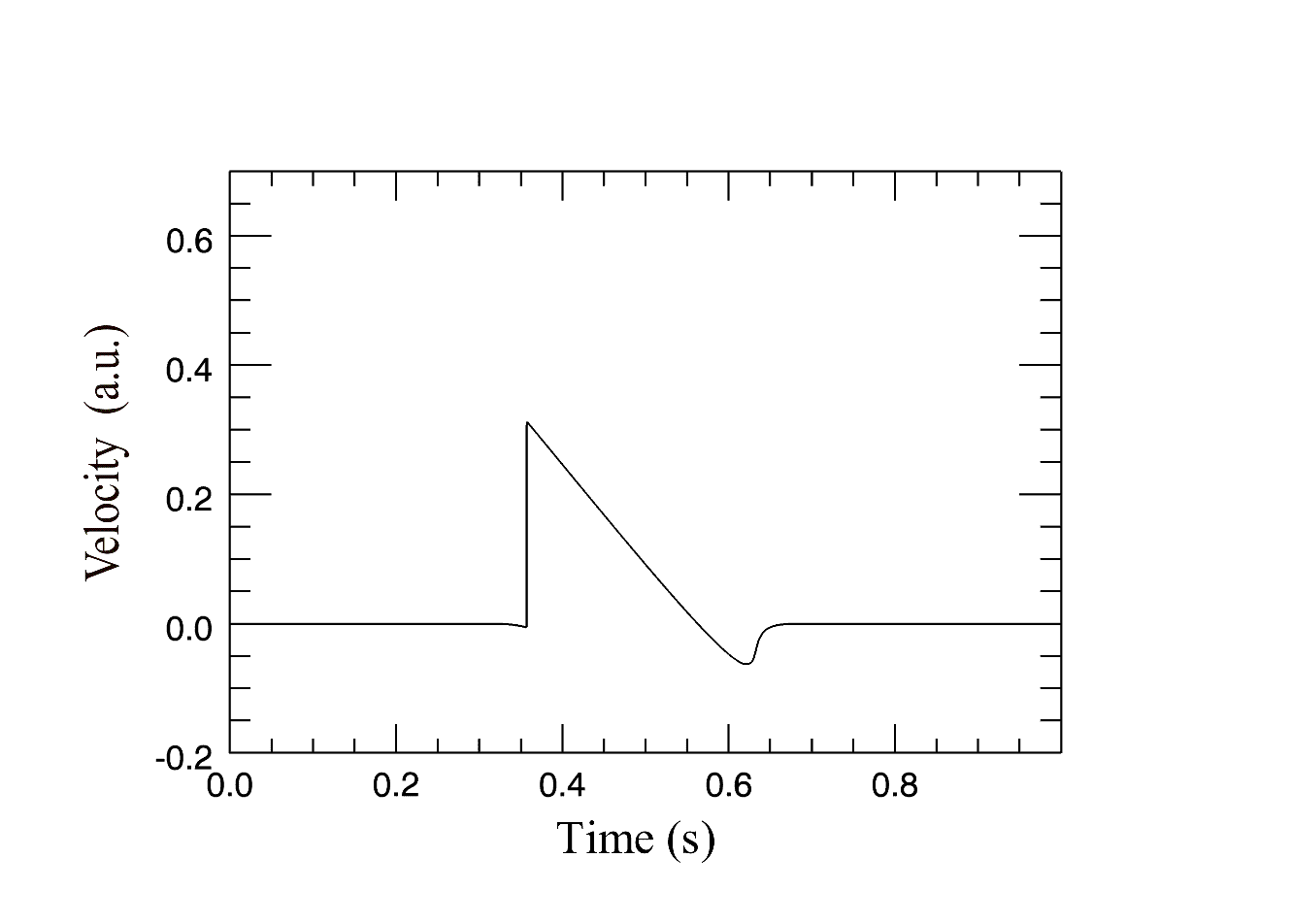}
\caption{The time development of the fully developed signal after 60.000 iterations. Video (showing steps of 5.000 iterations) available at https://doi.org/10.11583/DTU.12016881.v1 }
\label{fig:7}       
\end{figure}

\begin{figure}
  \includegraphics[width=\linewidth]{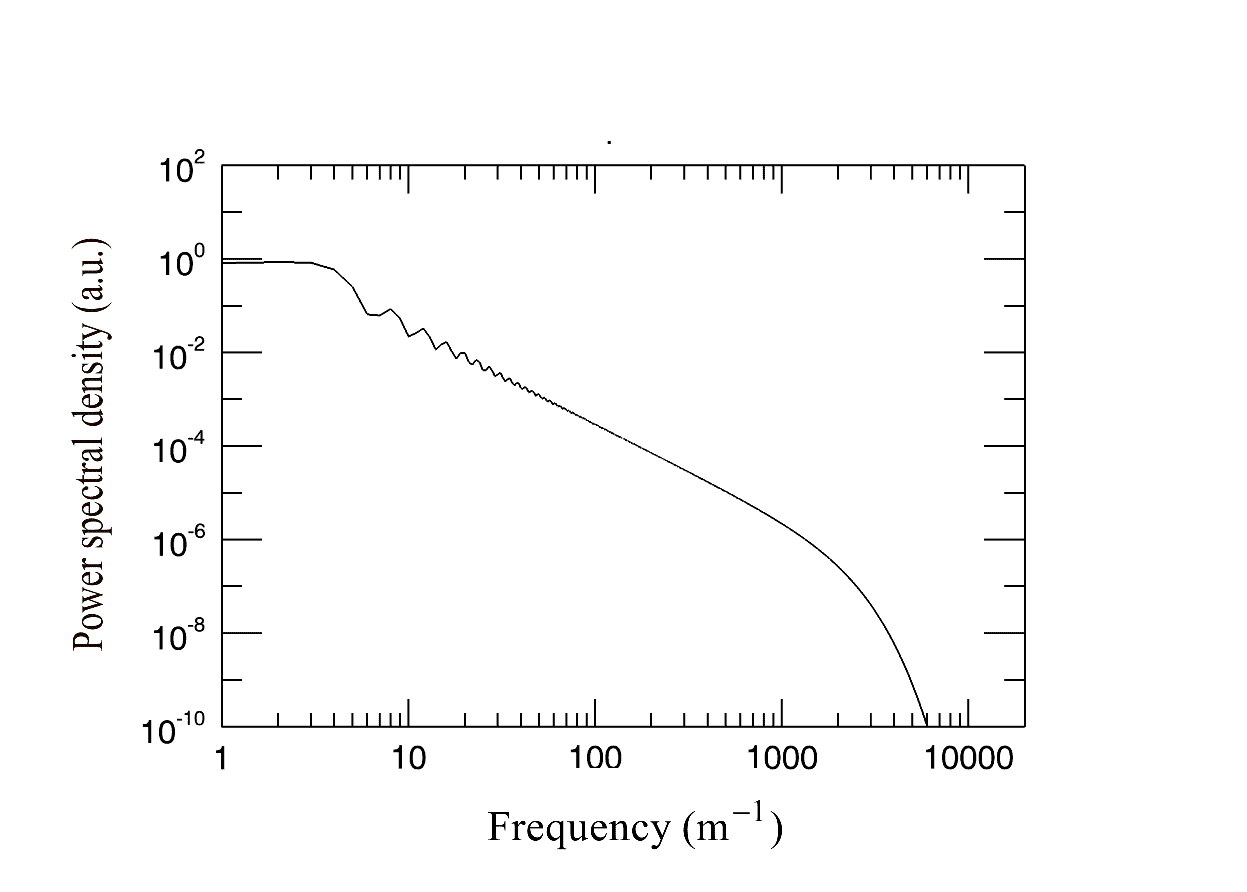}
\caption{The fully developed power spectrum after 60.000 iterations. Video (showing steps of 5.000 iterations) available at https://doi.org/10.11583/DTU.12016833.v1 }
\label{fig:8}       
\end{figure}

Figure~\ref{fig:7} shows a greater range of time development (movie showing steps of 5.000 iterations up to a total of 60.000 iterations, five times greater than in Figure~\ref{fig:5}). In this sequence, we note the following: The pulse develops further toward a shock-like form. However, a full development to a shock wave is prevented by the removal of high frequency energy by the dissipation. The sudden change in velocity across the pulse, e.g. in Figure~\ref{fig:7}, is manifested in higher frequency content in the associated power spectrum in Figure~\ref{fig:8}. As can also be seen in the wave-based formulation (Galerkin projection) of the turbulent kinetic energy equation in~\cite{10,11}, as well as from the definition of dissipation, the dissipation is associated with large values of the spatial gradients of the velocity waves. Considering that the steep gradients are predominantly concentrated around the shocks, it appears that the non-linear term in the Navier-Stokes equation causes intermittency in the dissipation of turbulent kinetic energy.

The power spectrum eventually develops to a fully developed form, which does not change further with time (Figure~\ref{fig:8}, movie showing steps of 5.000 up to a total of 60.000 iterations). Correspondingly, the velocity trace also develops into a fully developed, soliton like state. However, the spectrum energy is dissipated, and the spectrum is seen to decrease slowly in power, while the shape remains unchanged, an indication of a constant (or equilibrium) cascade in the inertial range.

\begin{figure}
  \includegraphics[width=\linewidth]{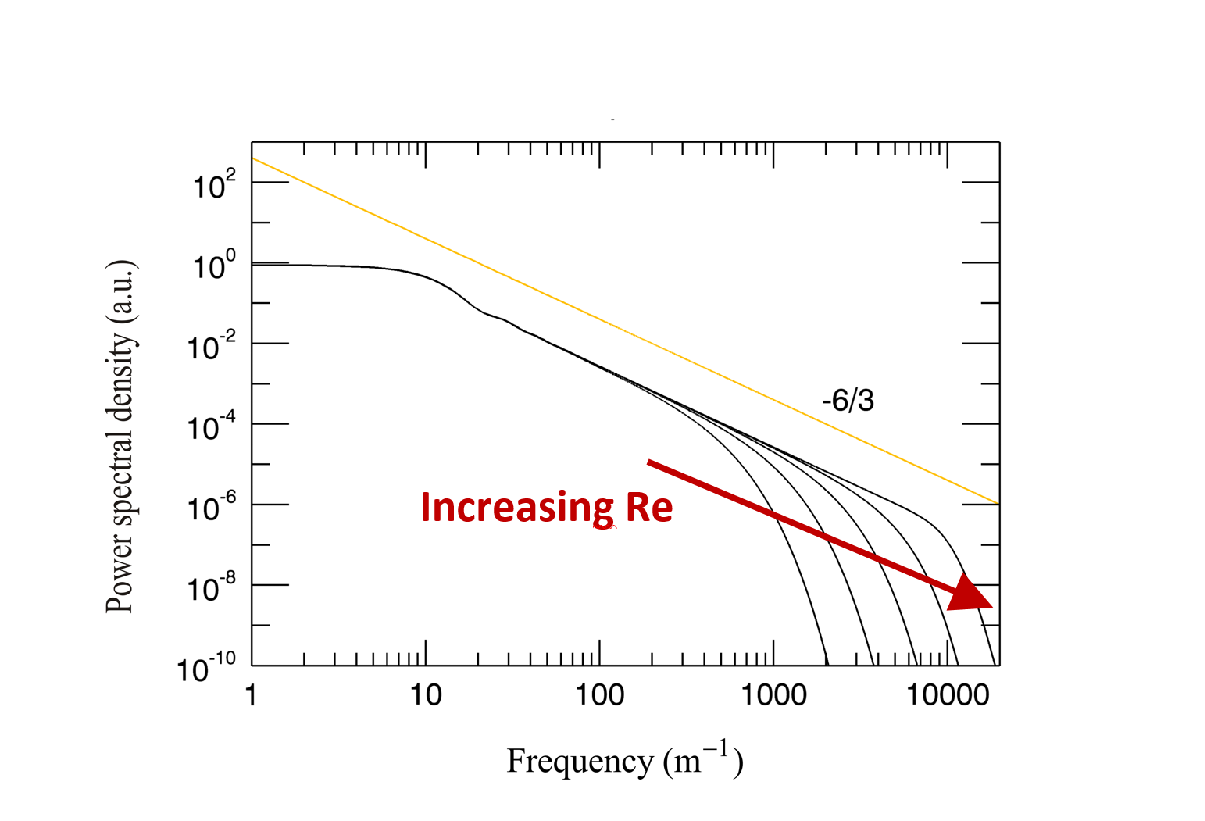}
\caption{Fully developed power spectrum for different values of the Reynolds number; $Re = 8$, $4$, $2$, $1$, $0.5$ $\cdot 10^{-3}$, respectively. Color figure available online. }
\label{fig:9}       
\end{figure}

Again, the power law of the power spectrum is cut off at high frequency due to dissipation. The high frequency cut-off of the power spectrum clearly depends on the Reynolds number, $Re = UL/\nu$, where $U=0.5\,ms^{-1}$ and $L= 0.25\, m$ are characteristic large-scale dimensions for the flow and the viscosity $\nu$ has been set to $8$, $4$, $2$, $1$, $0.5$ $\cdot 10^{-3}$ in Figure~\ref{fig:9}. When the Reynolds number, $Re \rightarrow \infty$, the situation has often been considered a paradox~\cite{11a,11b,11c}: How can the energy be dissipated when the viscosity $\nu \rightarrow 0$? The answer is that the energy is not dissipated in the limit where $Re \rightarrow \infty$, but the power spectrum is spread over ``an infinitely large'' spectral range for the same input energy but increasing Reynolds numbers (or decreasing kinematic viscosity $\nu$). The spectrum can, in principle, keep spreading to higher frequencies or wavenumbers without upper bound, as the viscosity is decreasing until we reach the limit for the continuum hypothesis.

\subsection{Time trace and velocity power spectrum of initial low Reynolds-number van K\'{a}rm\'{a}n-type signal}

\begin{figure}
  \includegraphics[width=\linewidth]{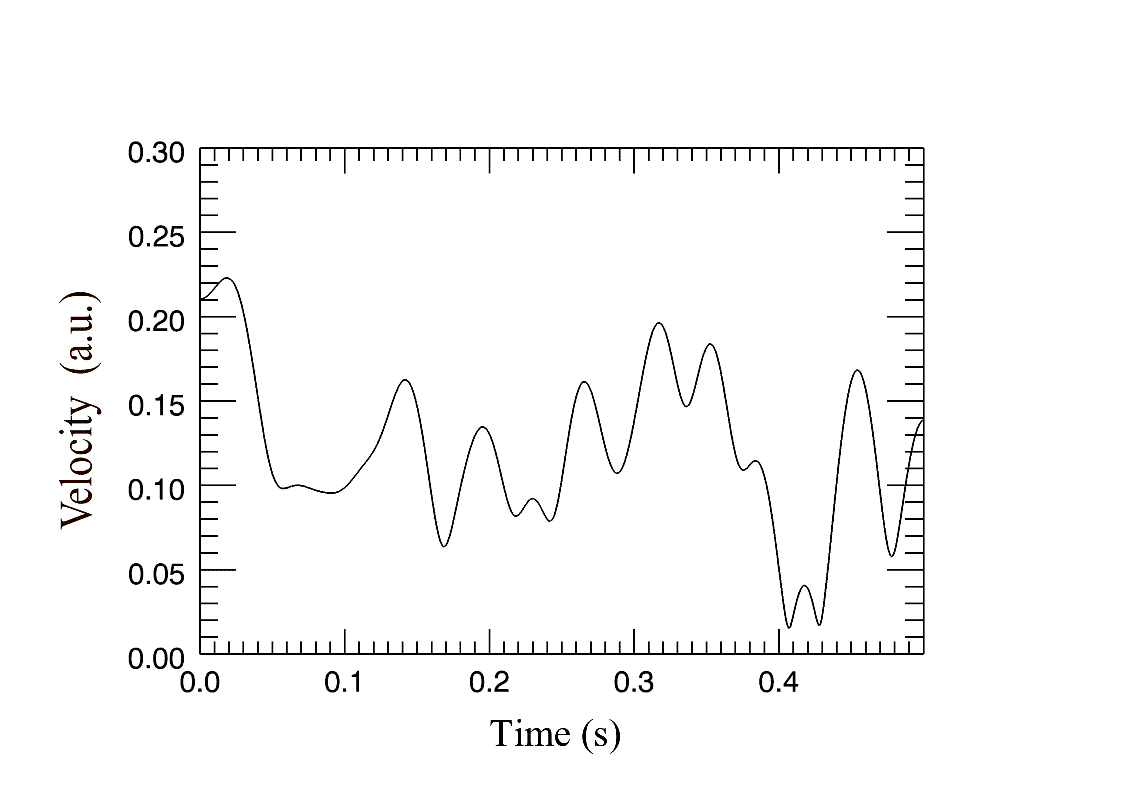}
\caption{Initial time trace with low frequency van K\'{a}rm\'{a}n spectrum.}
\label{fig:10}       
\end{figure}

\begin{figure}
  \includegraphics[width=\linewidth]{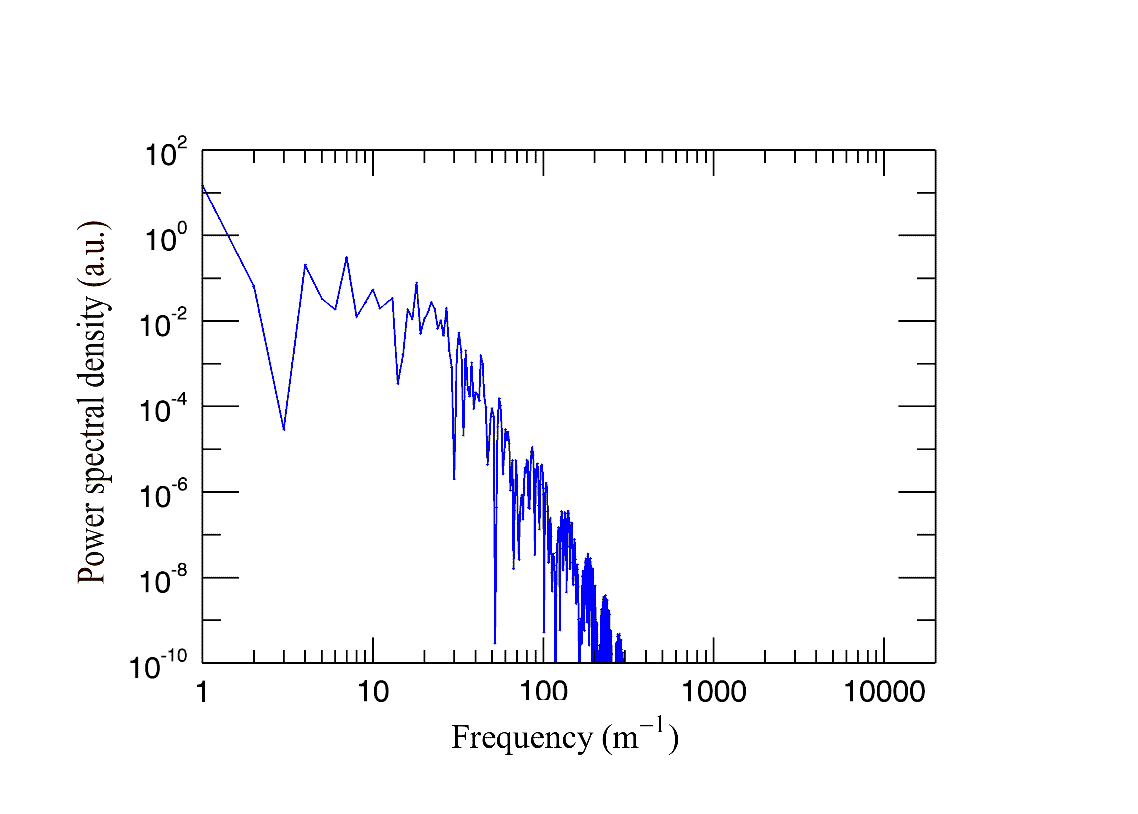}
\caption{Initial low frequency van K\'{a}rm\'{a}n power spectrum. Color figure available online. }
\label{fig:11}       
\end{figure}

\begin{figure}
  \includegraphics[width=\linewidth]{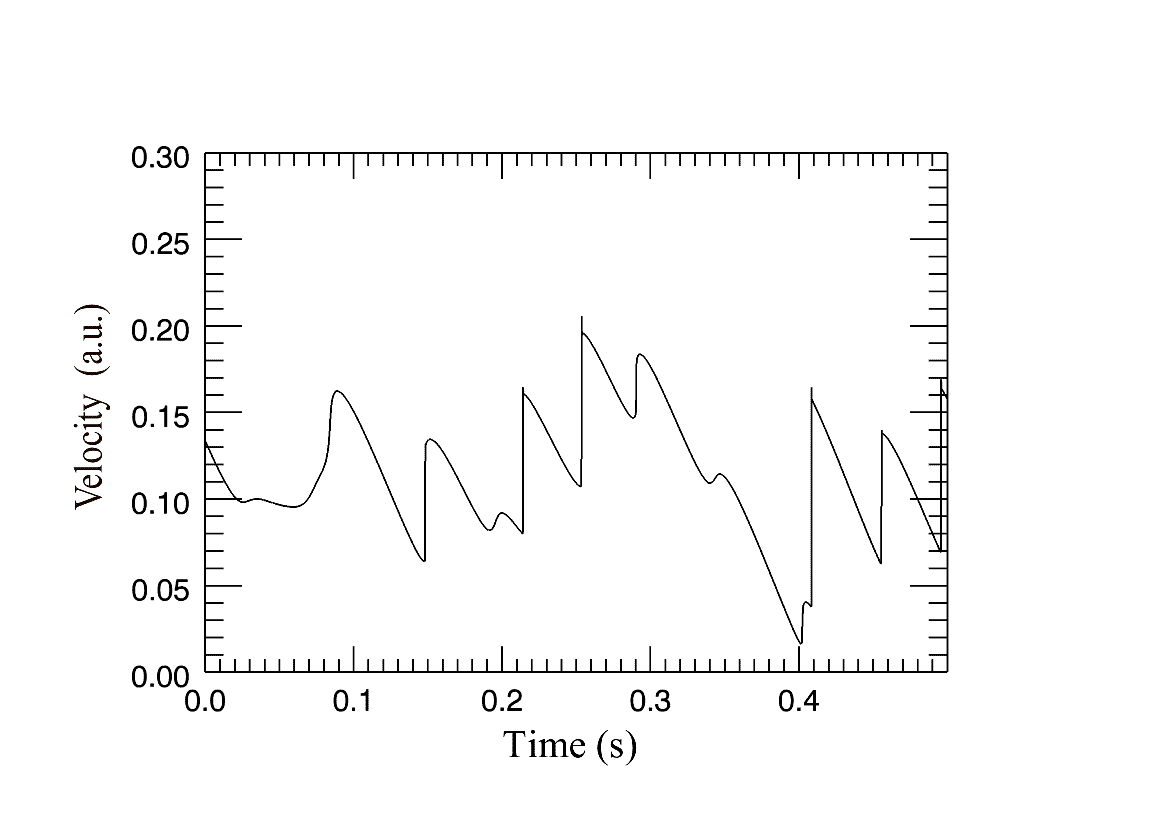}
\caption{Fully developed time trace with a van K\'{a}rm\'{a}n spectrum. Video available at https://doi.org/10.11583/DTU.12016860.v1 }
\label{fig:12}       
\end{figure}

\begin{figure}
  \includegraphics[width=\linewidth]{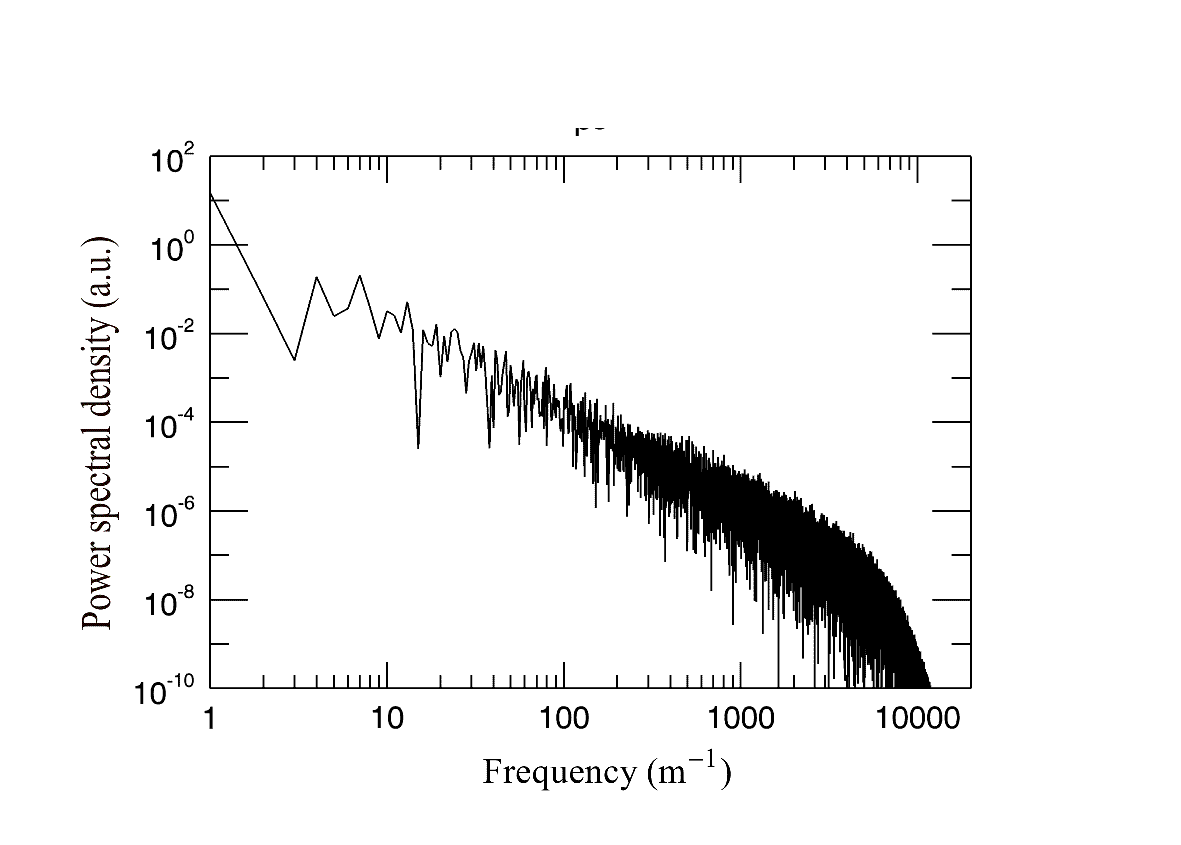}
\caption{Time development of van K\'{a}rm\'{a}n power spectrum. Video available at https://doi.org/10.11583/DTU.12016848.v1 }
\label{fig:13}       
\end{figure}

In this example we generate an incident velocity time record with a relatively low frequency van K\'{a}rm\'{a}n-type spectrum. Figure~\ref{fig:10} shows a single $0.5\,s$ time record and Figure~\ref{fig:11} shows the input power spectrum. Figure~\ref{fig:12} shows the fully developed time signal after $10.000$ iterations and Figure~\ref{fig:13} shows, correspondingly, how the velocity power spectrum has developed after $10.000$ iterations. Note, in Figure~\ref{fig:11} and~\ref{fig:13}, the power spectrum is calculated from a single realization corresponding to the time trace in Figure~\ref{fig:10} and~\ref{fig:12}. As can be seen from both the time trace and the corresponding spectrum (including in the previous examples), the large scales are largely unaffected with the flow development. This has historically been referred to as a `permanence of large eddies', or an ability of the flow to remember its initial conditions. The smaller scales develop with the repeated actions of the nonlinear term to yield steeper slopes in the time trace and, correspondingly, higher frequencies in the spectrum.

As the program lets the velocity record propagate through the CV multiple times, the time record and the corresponding power spectrum change shape as seen in the movies (see links in the figure captions).

Leaving out the pressure term (which in this context acts as a noise term) brings out in a clear way some of the effects of the nonlinear convection term as also seen in Figure~\ref{fig:7} and Figure~\ref{fig:8}. As the velocity record and the corresponding power spectrum are allowed to develop fully, they attain, once again, a final form that does not change further; one may speak of a soliton-like behavior. The only effect of further iterations is a slow reduction of the kinetic energy brought about by the dissipation term. This corresponds to a situation where the energy is provided at the outset as the initial velocity record and does not receive further contributions (decaying turbulence). Alternatively, one could have chosen to update the energy with a small amount for each iteration, which would eventually lead to a constant condition corresponding to forced turbulence. In the current example, however, it is clear that also the large-scale features of the time traces in Figures~\ref{fig:10} and~\ref{fig:12} are not altered significantly with time, even from the initial development. The results thus indicate that initial conditions are remembered by the flow even after significant development of the spectrum -- at least if left unforced.

The properties of the nonlinear response of the Navier-Stokes equation, which is brought out in the simple example of a deterministic pulse, is harder to observe in a turbulent flow because of the randomness of the velocity signal from the globally acting pressure term. Further randomness is introduced into the simulations when the pressure term is included. The pressure fluctuations are propagated through the CV so that the fluctuating pressure gradient across the CV from the integrated contributions of the surrounding flow adds random momentum (and random kinetic energy) to the flow.

\begin{figure}
  \includegraphics[width=\linewidth]{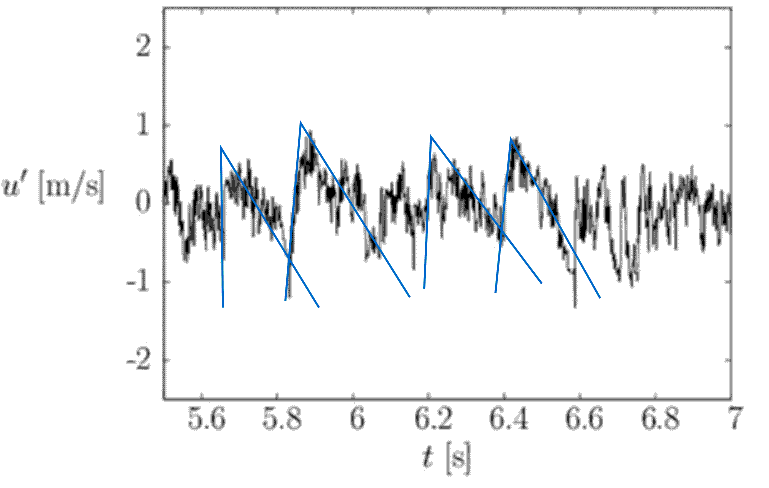}
\caption{Turbulent velocity of a parameterizable constant temperature anemometer (reproduced with permission) from Ndoye et al.~\cite{12}. A few blue lines have been overlaid to guide the eye. Color figure available online. }
\label{fig:14}       
\end{figure}

It is thus interesting to note that even measured signals display the same characteristic `sawtooth'-traits as described above (when one knows what to look for!), although somewhat obscured by the effect of turbulence production and pressure fluctuations. Figure~\ref{fig:14} (from~\cite{12}) shows the time trace of a hotwire measurement in a turbulent boundary layer at a wall normal distance (in wall units) of $y^{+}=26$. This behavior can be seen even in ``classic pictures'' of turbulent velocity records, e.g., Pope ``Turbulent Flows'' Figure~1.3~\cite{13} or Figure~2 in Sreenivasan~\cite{14}, and tends to be clearer in fully developed, unforced turbulence.

\subsection{Effect of pressure fluctuations, modeled pressure term}\label{sec:pressure}

The pressure term in the Navier-Stokes equation describes the total effect of pressure fluctuations generated elsewhere in the fluid or imposed at the boundaries. Local pressure fluctuations from the flow field is caused by velocity fluctuations (dynamic pressure) throughout the whole flow field, and these pressure fluctuations create pressure gradients in the CV. If the complete three-dimensional velocity field is known at the time, the local pressure is found by solution of a Laplace equation. The one-dimensional convection velocity through the CV thus cannot deliver the pressure from a 3D flow field. If we want to investigate how the pressure affects the development of the velocity record and the statistical quantities such as velocity power spectrum and structure functions, we are left to infer the pressure gradient term from modeling (c.f.,~\cite{15,16}), experimental data or assumptions. Theory and DNS computations can provide some suggestions, but the answers are complex, depending on Reynolds number and boundary conditions.

In the spirit of our attempts to provide simple solutions and results that help understand the basic physical processes governing the energy exchange in the turbulence, we propose a simple model: Assuming a homogeneous velocity field, where the velocity fluctuations at all points in the fluid have the same statistical properties as the velocity record describing the velocity in the CV, we can compute the dynamic pressure at all points of the flow. Adding statistically independent contributions from velocities all over the flow field, $\hat{u}_e(k)$, we find an expression for the pressure gradient in Fourier space:
\begin{equation}
\nabla \hat{p}(k)_p = \lambda \sum_{n=1}^{N} k_n \hat{u}_{e}(k_n)_{p}
\end{equation}

\begin{figure}
  \includegraphics[width=\linewidth]{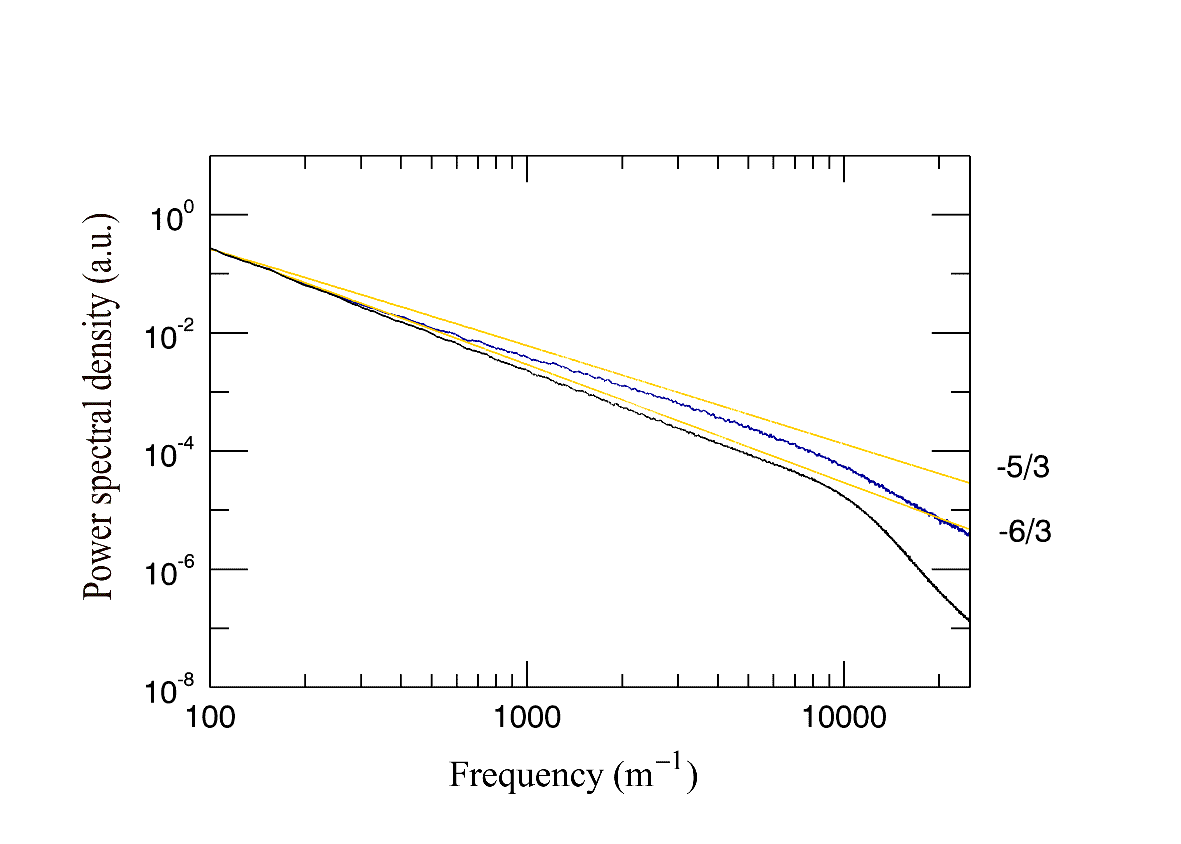}
\caption{Velocity power spectrum with and without pressure fluctuations. Color figure available online. }
\label{fig:15}       
\end{figure}

Without access to a theoretical expression or a DNS calculation, we cannot derive the magnitude of this contribution to the momentum fluctuations, which explains the adjustable parameter, $\lambda$. The effect of this term in the Navier-Stokes equation is to add to the momentum fluctuations and the power spectrum. The addition to the velocity power spectrum is a first order power law with a slope depending on $\lambda$.

Figure~\ref{fig:15} shows two fully developed van K\'{a}rm\'{a}n velocity power spectra originating from a low frequency, large eddy turbulence. The spectrum with the $-2$ (or $-6/3$) slope is from the Navier-Stokes equation without the pressure term. The spectrum with the $-5/3$ slope is the velocity power spectrum when pressure fluctuations resulting from statistically independent velocity fluctuations in an assumed homogeneous velocity field are included in the Navier-Stokes equation. The slope $-6/3$ is expected from the sawtooth-like form of the fully developed velocity record as explained above. The slope $-5/3$ results from the addition pressure variations from statistically independent velocity fluctuations with a power spectrum proportional to the wavenumber $k$. The parameter $\lambda$ has been adjusted to raise the slope to $-5/3$.

\subsection{Single Fourier mode injection}\label{sec:singleFmodeInjection}

To further study the interaction between Fourier modes, we have conducted a series of measurements and corresponding calculations where a single Fourier mode (a narrow-band signal) is injected into a well-known turbulent flow. Among others, the following two cases were studied~\cite{7}, which will be further analyzed herein:
\begin{itemize}
  \item \textbf{Case 1:} Measurement with a hotwire anemometer on a fully turbulent jet flow into which a Fourier mode was injected by means of an oscillating wing profile.
  \item \textbf{Case 2:} Measurement of a single Fourier component injected into a large low turbulence intensity jet core by vortex shedding from a rectangular rod.
\end{itemize}
In both cases, good agreement was found between measurements and computer simulations, validating the simulation method.\\

\begin{figure}
  \includegraphics[width=\linewidth]{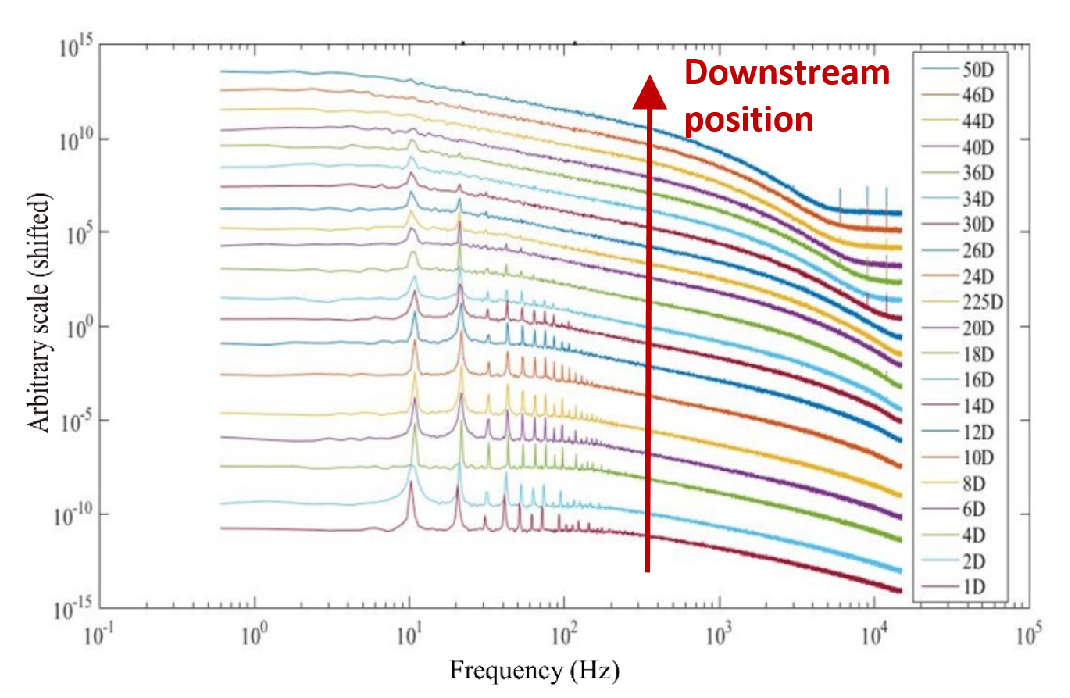}
\caption{\noindent Case 1: Velocity power spectra at increasing downstream distance from an oscillating wing in a fully developed turbulent round jet flow. Color figure available online. }
\label{fig:16}       
\end{figure}

\noindent \textbf{Case 1:} Figure~\ref{fig:16} shows the development of the power spectrum measured with hotwire anemometry at increasing distances from an oscillating wing positioned in a turbulent round jet flow. The wing was located $10$ jet exit diameters, $10D$, downstream from the exit of a turbulent jet. Note that for the sake of readability and comparison, the spectra have been shifted upwards along the ordinate with increasing distance from the wing profile trailing edge, with the distance indicated in terms of jet exit diameters, $D$.

It is noteworthy that the initial frequency has developed several higher harmonics even before reaching the trailing edge of the oscillating wing and stably retains these peaks across the downstream direction even in this highly turbulent flow. The shape of the spectrum of these harmonics is observed to quickly develop a distribution of energy between the peaks that does not change significantly further downstream apart from being eventually submerged in the background turbulence. This is well in line with the observation in the low Reynolds number van K\'{a}rm\'{a}n signal example that the time trace retains the large-scale features (`permanence of large eddies') and eventually develops into a soliton like decaying state.

\begin{figure}
  \includegraphics[width=\linewidth]{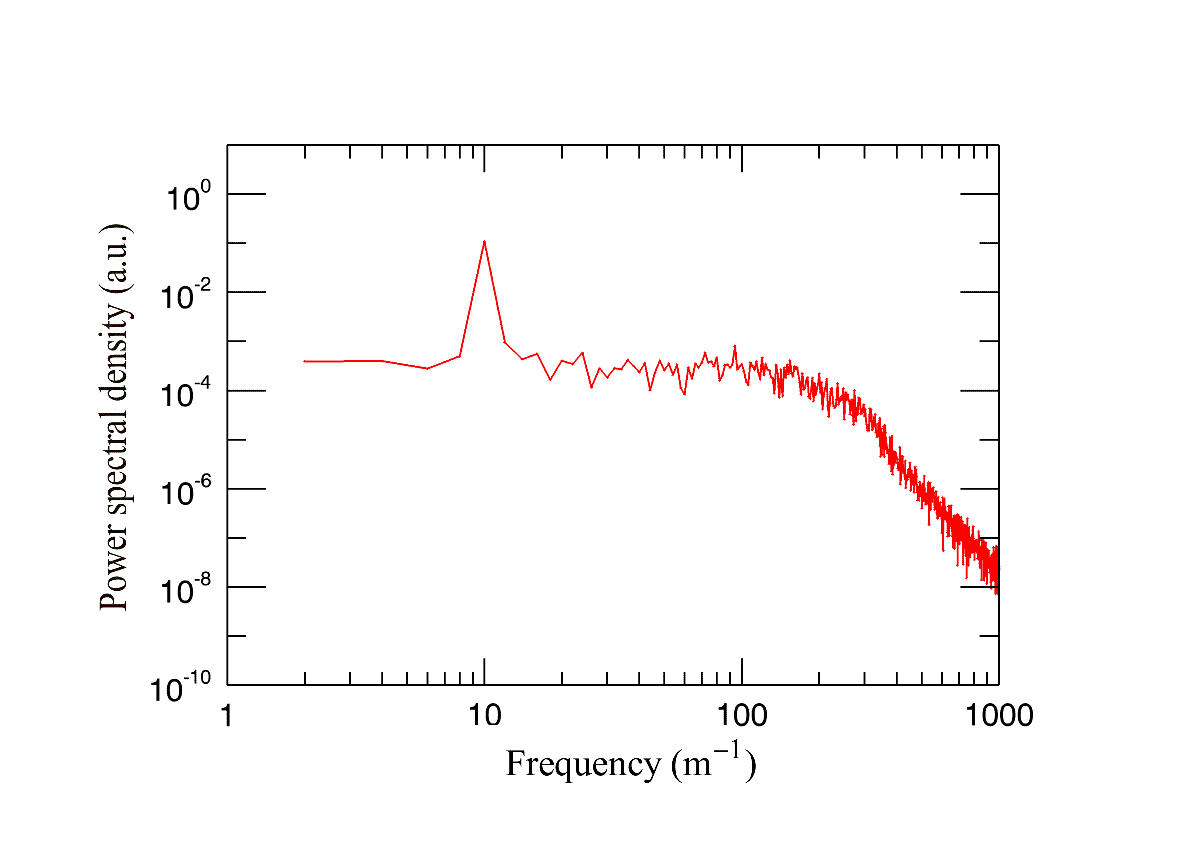}
\caption{Case 1: Simulated initial power spectrum of van K\'{a}rm\'{a}n type signal with superimposed narrow-band Gaussian pulse. Color figure available online. }
\label{fig:17}       
\end{figure}

\begin{figure}
  \includegraphics[width=\linewidth]{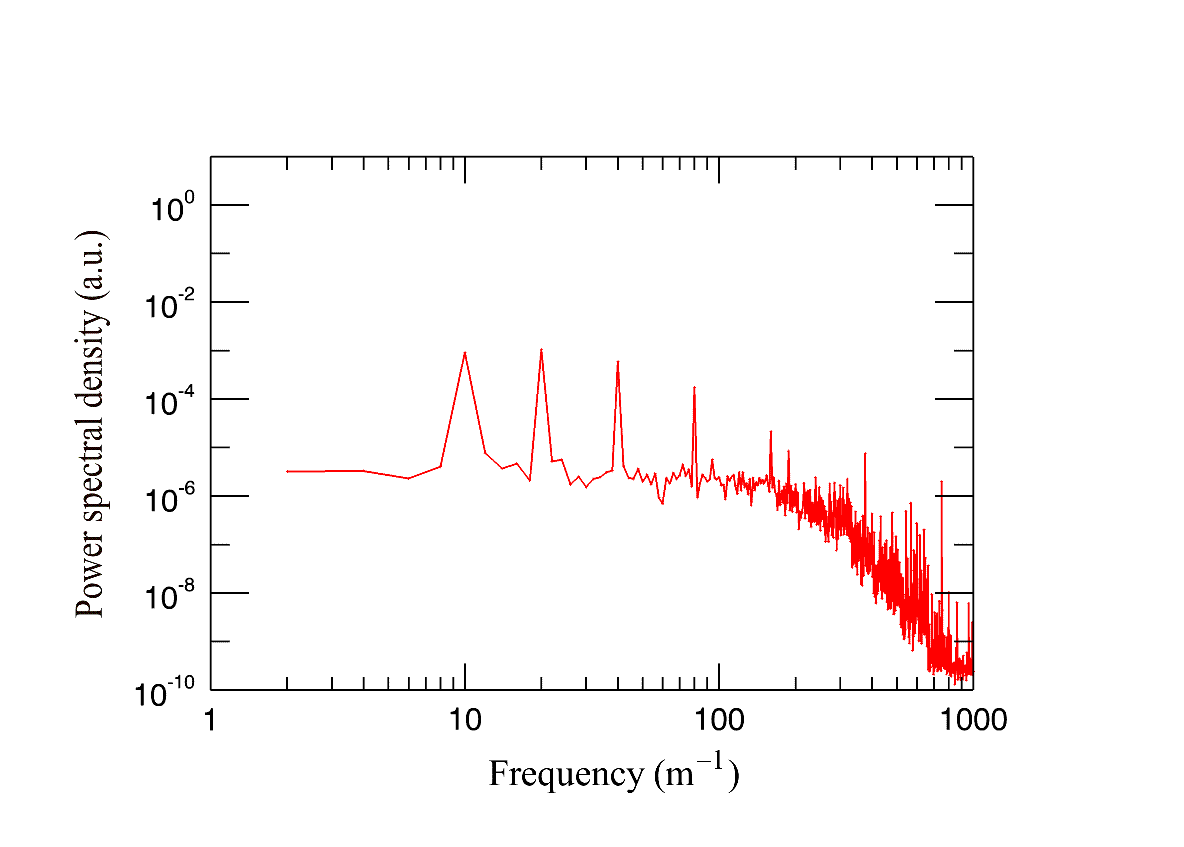}
\caption{Case 1: Simulated fully developed power spectrum. Color figure available online. }
\label{fig:18}       
\end{figure}

For comparison, Figure~\ref{fig:17} shows the initial power spectrum of a simulated velocity signal with high intensity turbulence; a stochastic signal with a van K\'{a}rm\'{a}n spectrum superimposed with a narrow band Gaussian pulse. Figure~\ref{fig:18} shows the fully developed power spectrum after $20.000$ iterations, simulating repeated actions of the Navier-Stokes equation corresponding to a time development of the signal and the power spectrum. Both the measured and computed spectra display the time development of the large mode injected at the outset, which allows determination of the time delay in the cascade process. Also apparent is the relatively low efficiency for the creation of the third harmonic mode and subsequent higher odd order modes, as also observed in the measurements in Figure~\ref{fig:16}. In the specific simulation result in Figure~\ref{fig:18}, these odd harmonics are in fact suppressed down to the statistical noise level of the spectrum. This effect is attributed to the influence of the spatial windowing caused by the finite size of the flow region, an effect that is described in detail in~\cite{8}.\\

\noindent \textbf{Case 2:} Another example illustrating the development of a narrow-band mode (symbolizing a single Fourier mode), is described in~\cite{7}. A regular sinusoidal velocity oscillation is created by vortex shedding behind a rectangular rod (with sharp corners) inserted across a relatively low intensity turbulent air jet core. The velocity is measured by hotwire anemometry as $1\,s$ long time records at regular positions downstream of the rod, and the power spectra are computed from the velocity records. 

\begin{figure}
  \includegraphics[width=\linewidth]{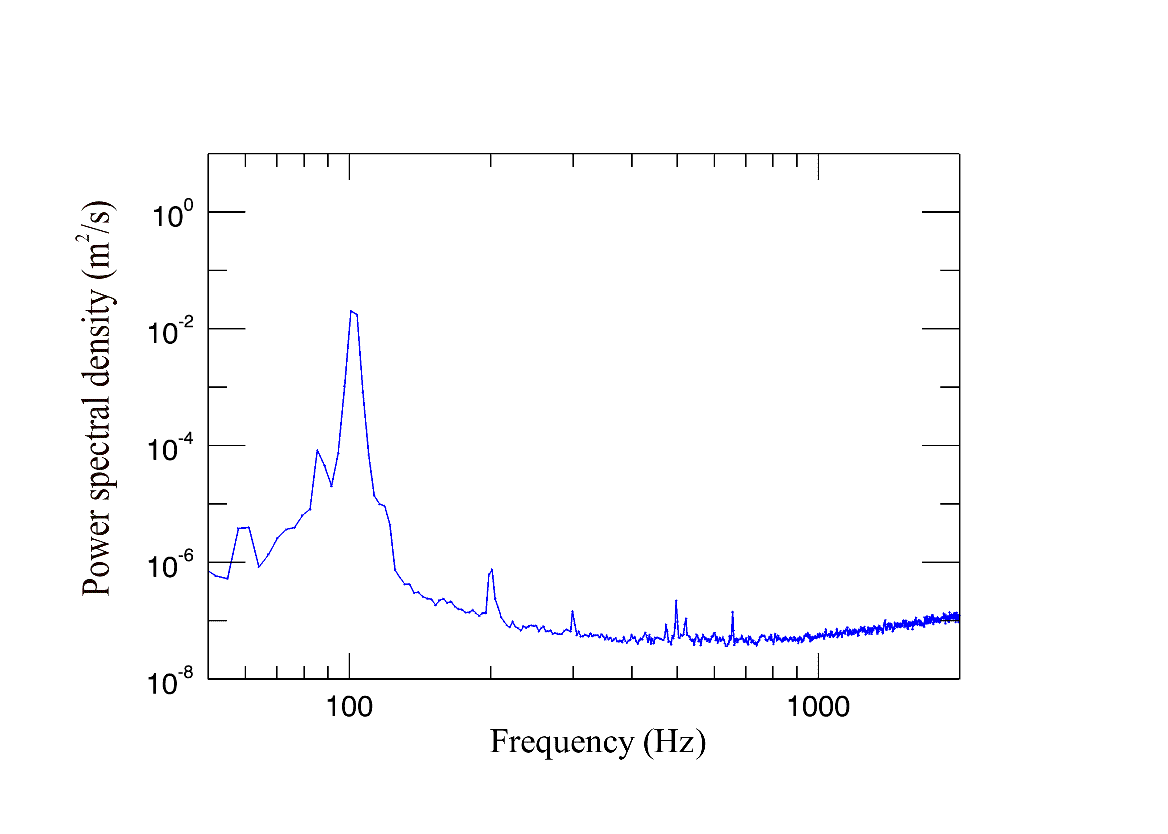}
\caption{Case 2: Measured vortex shedding $1\,mm$ downstream from rod. Color figure available online. }
\label{fig:19}       
\end{figure}

\begin{figure}
  \includegraphics[width=\linewidth]{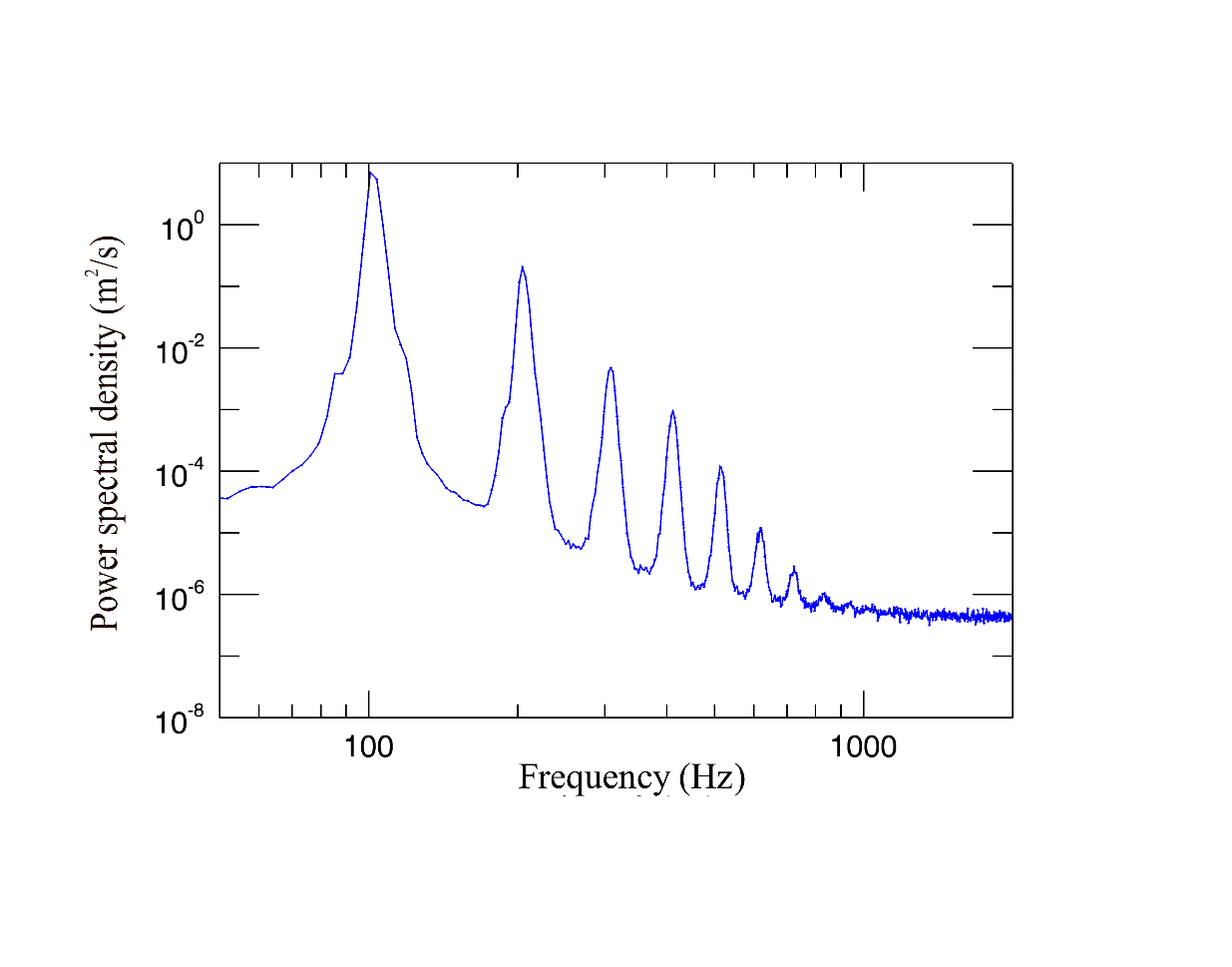}
\caption{Case 2: Measured vortex shedding $10\,mm$ downstream from rod. Color figure available online. Video available at https://doi.org/10.11583/DTU.12016827.v1 }
\label{fig:20}       
\end{figure}

Figure~\ref{fig:19} shows the initial power spectrum very near the rod and Figure~\ref{fig:20} shows the power spectrum at a position $10\,mm$ downstream. The move linked to Figure~\ref{fig:20} shows the power spectra at downstream positions in steps of $0.5\,mm$. Knowing the convection velocity, the spectral plots are interpreted as subsequent stages in the time development of the flow after being subjected to the effects of the Navier-Stokes equation during increasing time. The measurement reveals the action of the cascade and the time constants associated with the transfer of energy from mode to mode. The observed slow energy variations across frequency are a result of the default filter settings of the acquisition instrument~\cite{7}.

\begin{figure}
  \includegraphics[width=\linewidth]{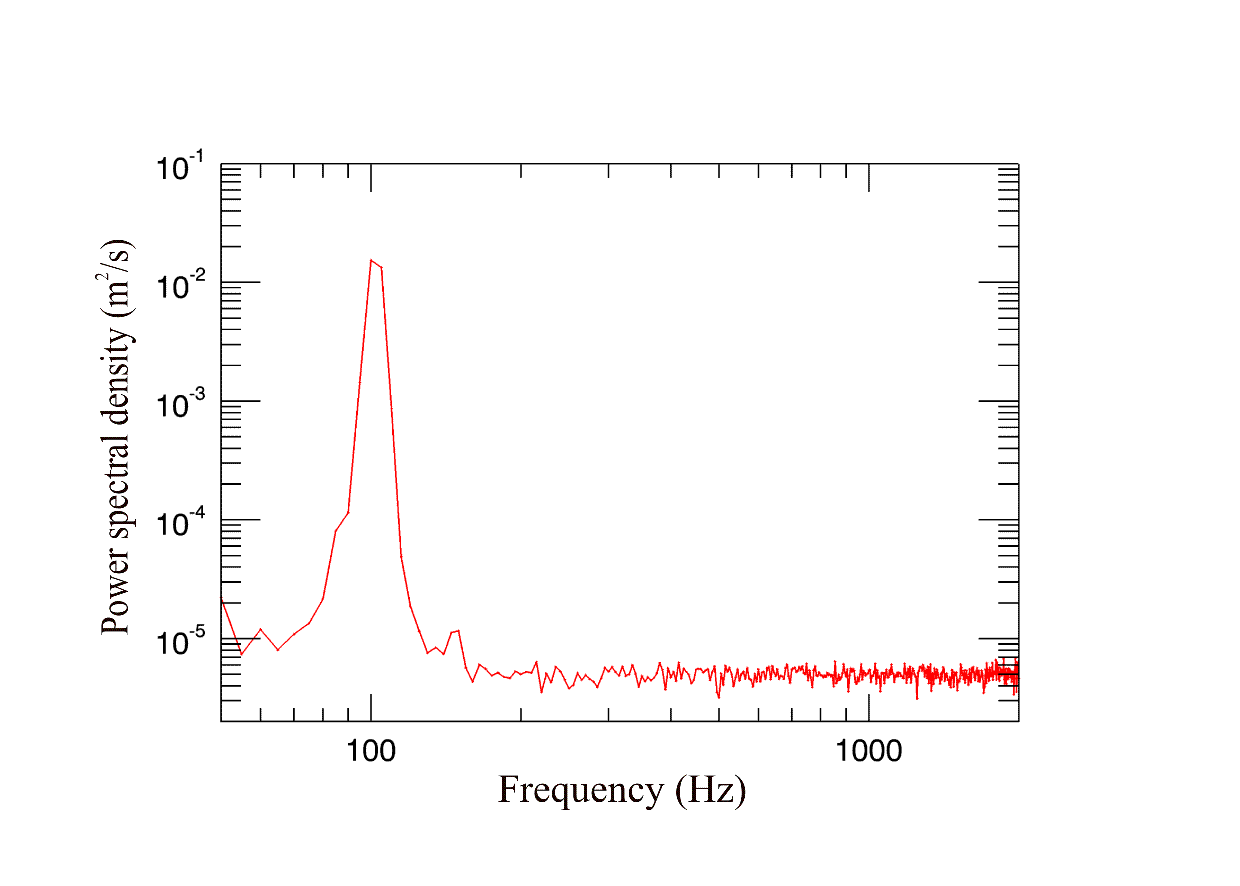}
\caption{Case 2: Simulation initial input from experiments of vortex shedding $1\,mm$ downstream from rod. Color figure available online. }
\label{fig:21}       
\end{figure}

\begin{figure}
  \includegraphics[width=\linewidth]{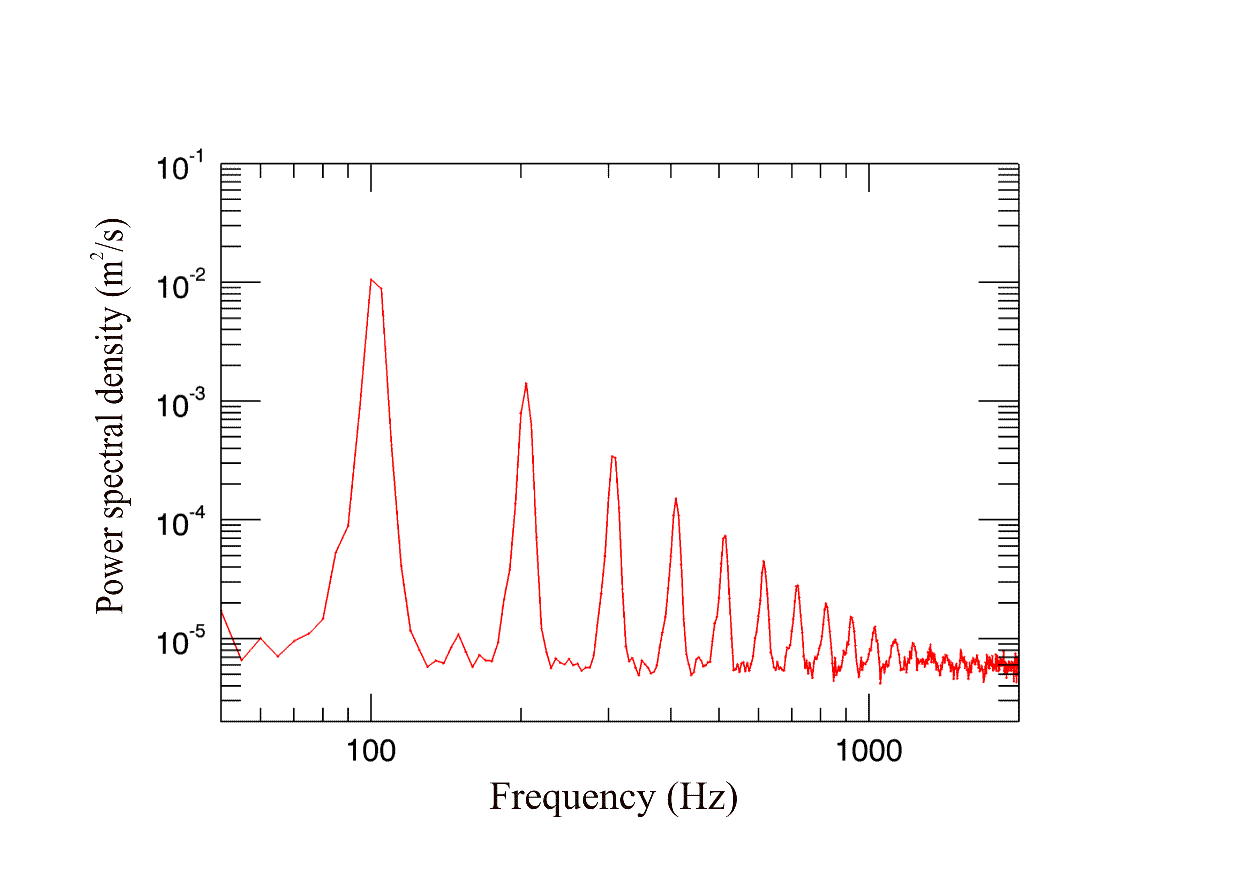}
\caption{Case 2: Simulated vortex shedding $10\,mm$ downstream from rod. Color figure available online. Video available at https://doi.org/10.11583/DTU.12016869.v1 }
\label{fig:22}       
\end{figure}

The flow development was subsequently simulated by computation with the Navier-Stokes Machine. To create a realistic comparison between experiment and computation, the $1\,s$ hotwire anemometry digital time record measured near the rod was simply used as input for the Navier-Stokes Machine calculation. Figure~\ref{fig:21} shows the initial time record as displayed by the program (the slight difference in shape between the spectra in Figure~\ref{fig:19} and Figure~\ref{fig:21} is due to differences in frequency response and display conditions). Figure~\ref{fig:22} shows the power spectrum after $100.000$ iterations. The movie linked to Figure~\ref{fig:22} shows the time development of the power spectrum. The similarity between the spectrum measured as a fluid element that is convected downstream from the initial creation of the vortex and the spectra computed by repeated small actions of the Navier-Stokes equation is striking and corroborates the computational method based on the development of the spatial convection record~\cite{6}.

\subsection{Non-local interactions}

\begin{figure*}
  \center{\includegraphics[width=0.9\textwidth]{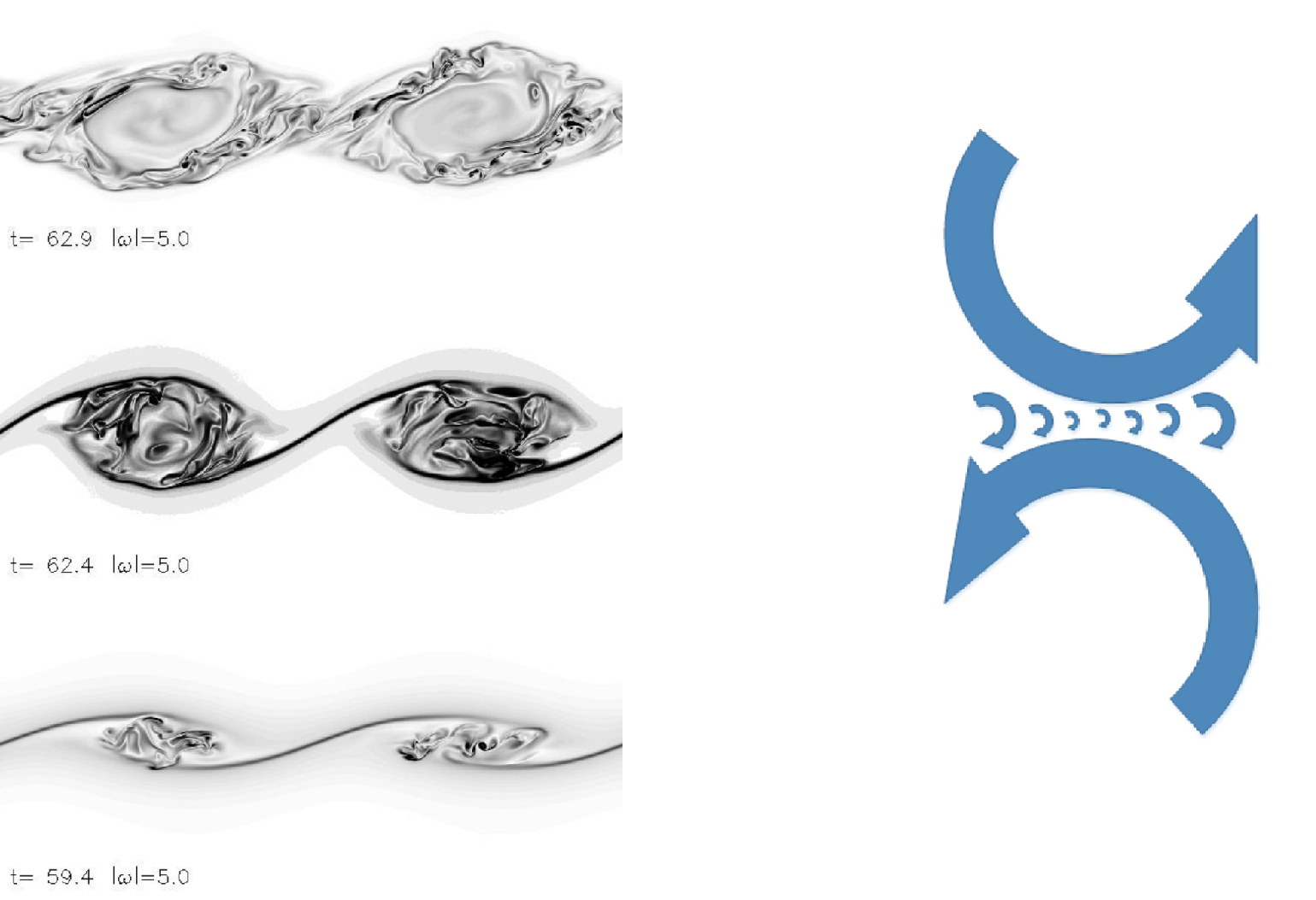}}
\caption{(Left) Example of shear layer simulation with nonlocal interactions~\cite{17} (reproduced with permission). (Right) Conceptual sketch of nonlocal interactions. Color figure available online. }
\label{fig:23}       
\end{figure*}

The manifestation of the sharp pulses of these single point time traces in a regular flow visualization setting is not trivial. The time records are a representation of the spatial structures corresponding to the fluid particle development at an Eulerian point. The spectra show, correspondingly, the frequency content of the time trace, or, if many realizations are averaged, the average frequency content of the flow at the point. We have observed that time traces at a point (as `seen' by e.g. a hotwire anemometer) display the generation of pulses with steepness bounded by the finite viscosity. Velocity waves, on the other hand, tend to `curl up' and thereby generate higher frequencies, as observed in a flow visualization, see e.g. Figure~\ref{fig:23} (left) for an example from a shear layer~\cite{17}. Thus, it appears reasonable that the sharp pulses are a manifestation of intermittency in the dissipation and that the intermittent dissipation in some manner is coupled to high shear.

A thought experiment may be helpful in hypothesizing qualitatively how large and small scales may couple together: Assume two large and closely spaced counter-rotating vortices, Figure~\ref{fig:23} (right), which can immediately be seen to give rise to the creation of significantly smaller scales. The large scales can in this manner exchange energy directly with the smaller scales and thus give rise to intermittent dissipation. This is an intuitive example in physical space of non-local interactions in wavenumber space, aiding direct energy transfer between widely different scales -- including non-equilibrium dynamics immediately transferred from large to small scales.

It is often emphasized that modal interactions do most efficiently occur between closely aligned modes in wavenumber space (local interactions) with a net energy transfer towards higher frequencies or wavenumbers. But another important aspect should be considered when discussing modal interaction: The time duration of a coherent interaction. The time issue is rarely considered, but it should be noted that if a large eddy interacts with a small one coherently over a long time, the transfer of energy between remote modes may still happen efficiently. This could typically happen e.g. in shear layers~\cite{10,11,17} (see also Figure~\ref{fig:23} left) or in fractal grid turbulence~\cite{18}, where different (in some cases quite distant in wavenumber) scales are introduced with a common convection velocity.

\begin{figure}
  \includegraphics[width=\linewidth]{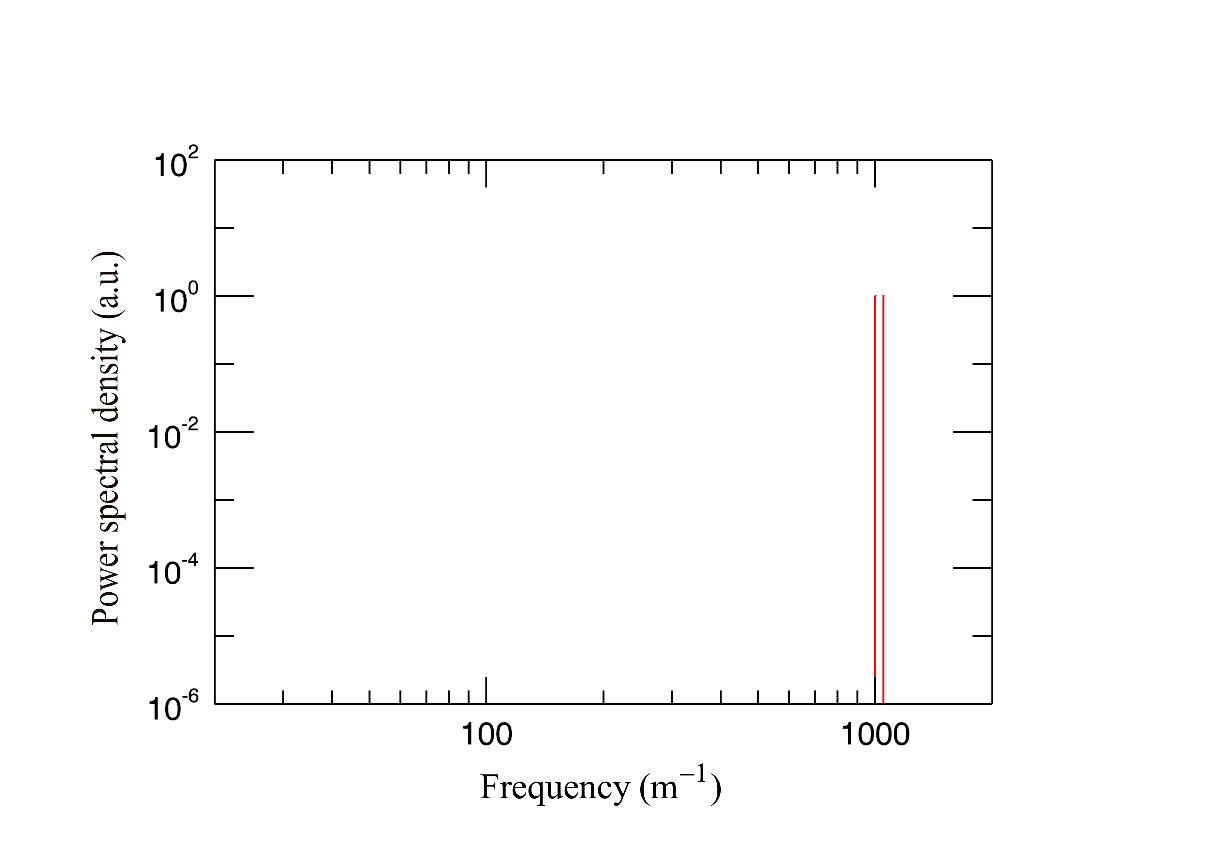}
\caption{Simulation: Two high frequency modes as the input to the simulation algorithm. Color figure available online. }
\label{fig:24}       
\end{figure}

We have illustrated the modal mixing over time with the following plots that show the effect of significant nonlocal interactions; In Figure~\ref{fig:24} two closely spaced high frequency modes create a low frequency mode far from the incident modes. Figure~\ref{fig:26} displays another situation where two modes separated by a large frequency distance interact over time to create closely spaced side bands.

\begin{figure}
  \includegraphics[width=\linewidth]{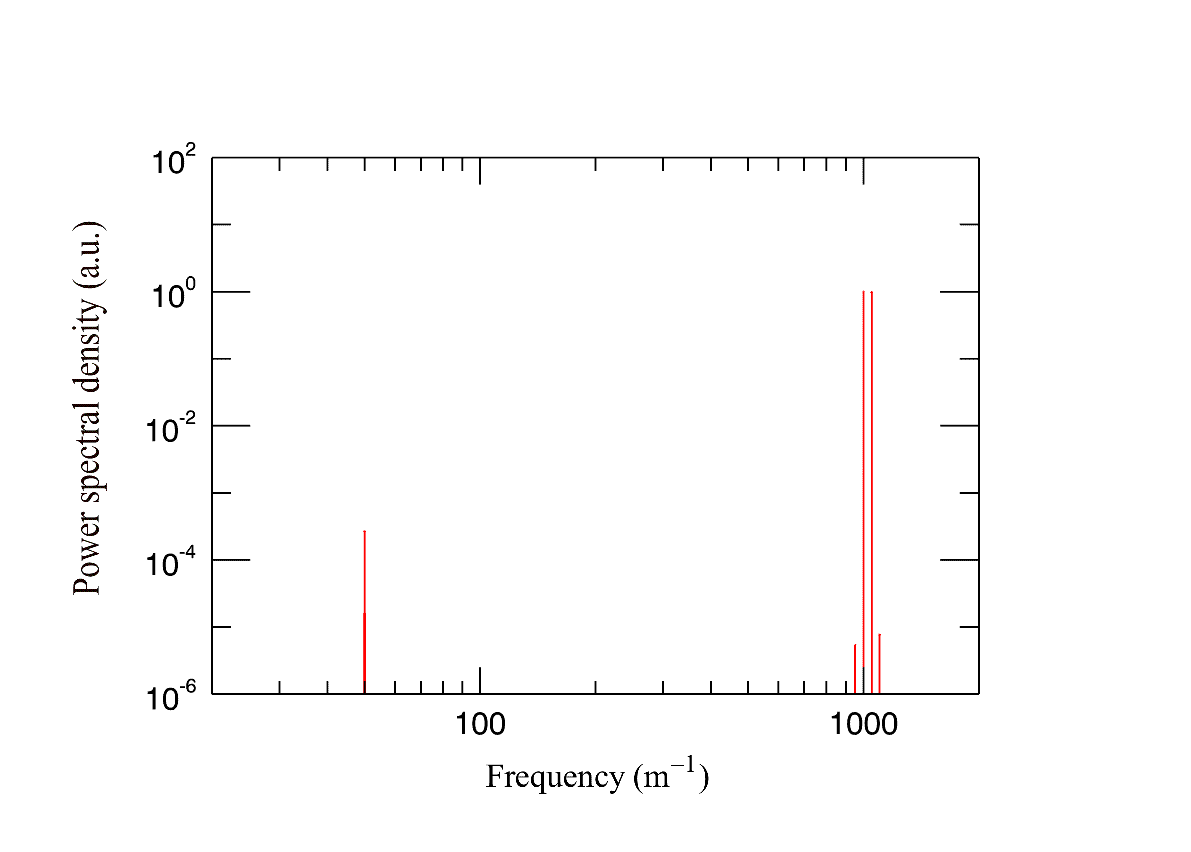}
\caption{Simulation: A low frequency mode generated far from the initial modes, where the initial modes are allowed to interact over time. Color figure available online. Video available at https://doi.org/10.11583/DTU.12016884.v1 }
\label{fig:25}       
\end{figure}

\begin{figure}
  \includegraphics[width=\linewidth]{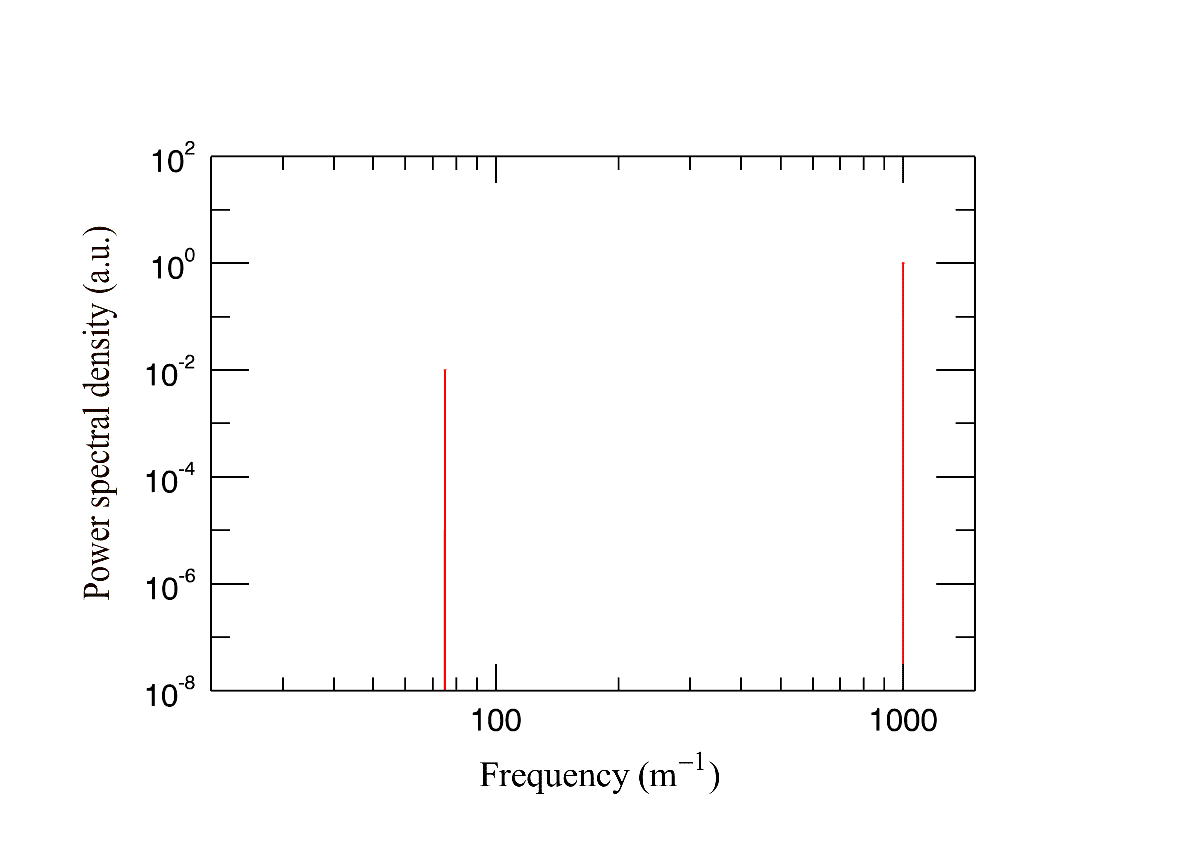}
\caption{Simulation: Two initial modes located far from each other, as the input to the simulation algorithm. Color figure available online. }
\label{fig:26}       
\end{figure}

\begin{figure}
  \includegraphics[width=\linewidth]{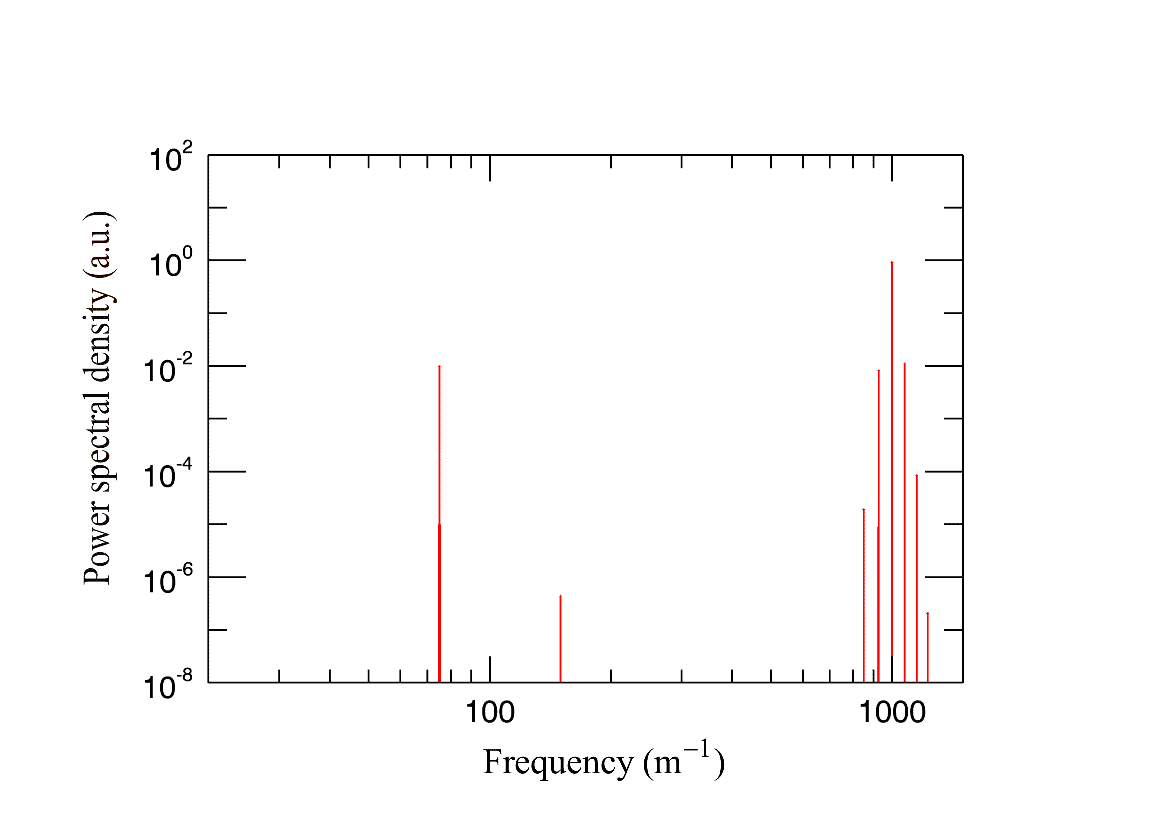}
\caption{Simulation: Closely spaced modes generated by nonlocal interaction, where the initial modes are allowed to interact over time. Color figure available online. Video available at https://doi.org/10.11583/DTU.12016821.v1}
\label{fig:27}       
\end{figure}

The movies linked to Figures~\ref{fig:25} and~\ref{fig:27} show how a low frequency mode can interact with high frequency modes over time to generate spatial frequency modes far from the energy carrying modes by non-local interactions. The movies also clearly show the tendency for preferred transfer towards higher frequencies, the energy cascade process, as was previously shown using propagation of two frequencies in the nonlinear term $u\cdot \nabla u$ (see section~\ref{sec:GaussianPulse}). It is thus the product of the velocity with its own velocity gradient that gives rise to these nonlocal interaction effects. Such high velocities and spatial velocity gradients combined are typical of high shear flows, e.g. such as those shown in Figure~\ref{fig:23}. Significant time duration of these interactions contribute to further increase the nonlocal energy exchange across wavenumbers.

\section{\label{sec:DiscussionConclusions}Discussion and Conclusion}
Probably, the most important result of this study is that it allows us to see the effect of the nonlinear term in the Navier-Stokes equation and study details of the triad interactions on a velocity record from a real turbulent flow measured in the laboratory.

In agreement with our stated purpose, to use a simple iterative computer program to help illustrate and understand the consequences of the specific form of the nonlinear term in the Navier-Stokes equation, we have presented examples of the time development of some typical input signals representing some simple flow situations. We have used a method that allows us to follow the interaction of individual so-called triad interactions, interactions between pairs of spatial Fourier components, and clearly see their development in time, illustrated, for example, in the development of the shape of the velocity time record and in the time development of the corresponding power spectrum.

From a simple example of a Gaussian pulse in time containing a well-defined range of frequencies, we have seen how the cascade of frequency components eventually evolves into a fully developed form. The repeated actions of the nonlinear term in the Navier-Stokes equation causes the formation of skewed pulses, whose steepness is bounded by viscosity and is directly linked to the intermittency of dissipation. The formation of these skewed pulses can be qualitatively understood from the iterative development in time of the product between the signal, $u$, and its gradient, $\nabla u$, in the nonlinear term. When reaching an equilibrium state, the time trace and the spectrum retain their shapes, and the latter is only modified by a slow reduction of the total energy caused by dissipation (for the case of decaying turbulence). 

It is also clear how the spectrum stretches to higher frequencies as the Reynolds number increases when viscosity is reduced. It is observed that the spectrum can in principle keep spreading to higher frequencies or wavenumbers without upper bound, as the viscosity is approaching the zero limit.

We have recognized the appearance of these skewed pulses, even for a highly random (turbulent) velocity record simulated using a van K\'{a}rm\'{a}n spectrum. Again, we see how a final form is approached that depends on the detailed properties of the input record (corresponding to the boundary conditions for a three-dimensional flow).

The deterministic pulses formed by the non-linear term can be observed even in `classical' textbook illustrations~\cite{13,14} of measured signals where pressure mainly acts as a background noise adding further to the kinetic energy and power spectrum. The steepness of these pulses is directly linked to the existence of high frequency energy in the power spectrum representation of the time trace, and we clearly see how the power law is cut off at high frequency due to finite dissipation. Dissipation is directly associated with large values of gradients in physical space and time, which we observe is predominating concentrated around the steep gradients in the pulses. Considering that the steep gradients are highly concentrated around the pulses, it appears that their existence, stemming from the non-linear term in the Navier-Stokes equation, constitute a cause of intermittency in the dissipation of turbulent kinetic energy. The coupling of these pulses to the appearance of regions of high shear and nonlocal interactions across wavenumbers was discussed as well as the possibility for large scales to transfer nonequilibrium dynamics directly to small scales. Furthermore, the large-scale features of the unforced time trace are observed to be preserved throughout the flow development, ranging even into the soliton-like state of decay.

The convection record method facilitated the analysis of interactions between analytical wave components in the nonlinear term, $u\cdot \nabla u$. The simple example presented in section~\ref{sec:GaussianPulse} showed how the nonlinearity in the Navier-Stokes equation does indeed cause a net shift in energy towards higher frequencies or wavenumbers, due to the presence of the spatial velocity gradient in the expression. From the manner in which the terms were weighted by sum and difference frequencies, it was observed that local interactions are dominating in the nonlinearity of the Navier-Stokes equation. Further, it was shown how this second order non-linearity in the fully developed form gives rise to a $-2$ power law if the pressure is not included in the analysis. This was seen to arise directly from a Fourier transformation of a triangular function as in the triangular shape of the pulses in the time records. The pulses, giving rise to this power law in the signal frequency content, appear to be a real flow property (based on the Navier-Stokes equation) and not just an artifact of Burgers' equation.

In the first two simulation examples, the pressure gradient contribution was not included. The resulting spectra thus have a slope of $-2 \,=\, -6/3$. The missing pressure gradient contribution accounts for the missing $+1/3$ contribution to obtain the classically expected $-5/3$ slope in the spectrum. However, due to the elliptic nature of the pressure contribution in the Navier-Stokes equation, the exact resulting dynamic pressure contribution to the spectrum would be expected to vary with flow and boundary conditions. The ``Navier-Stokes Machine'' program is local in physical space, based on a one-dimensional projection onto the flow direction of the forces acting on the fluid in an infinitesimal control volume. Thus, computation of the pressure gradient term is outside the scope of this program, dynamic pressure being a global flow phenomenon. Instead we have used an example where we assumed a homogeneous flow, where pressure fluctuations were created by velocity fluctuations from velocity records having the same statistical properties as, but uncorrelated with, the flow in the control volume. Due to the first order derivative of the pressure, the kinetic energy added from the pressure fluctuations is seen to raise the $-6/3$ slope caused by the convection term to a power law with a higher slope, e.g. a $-5/3$ slope.

In subsequent sections we showed how a ``single Fourier component'', a narrow-band oscillation added to the flow, can be followed in time and how and when higher order Fourier components are created. It is striking how the initial large-scale structure is preserved downstream in the flow and how the added frequency components remain well defined downstream -- even in a highly turbulent flow. In one example, the input to the simulation was taken from an experiment, which measured the development of the velocity power spectrum at a number of downstream locations interpreted as repeated action of the Navier-Stokes equation on the fluid. The power spectra computed with the measured velocity record as input showed excellent agreement between measurement and simulations, indicating relevance of the time-step method.

Furthermore, the impact of time duration on the efficiency of the interactions was illustrated by significant interactions between nonlocal modes in wavenumber space. Furthermore, if they are allowed to interact during significant durations in time, structures of widely different wavenumbers can be seen to exchange significant amounts of energy.

It is of course relevant to ask how well these relatively simple calculations, carried out on a laptop PC, can represent the properties of a real, three-dimensional turbulent flow. We are not able to answer this question in a general way, but in the cases where we have been able to compare simulations to actual measurements, we have observed a striking agreement. In particular, where we have tried to use a real measured initial velocity record, sampled from a laboratory flow experiment, and used that as input to the computer program, we have seen convincing agreement well downstream from the initial measurement point between measured and computed velocity power spectra.


\section*{Acknowledgements}
CMV acknowledges financial support from the European Research Council: This project has received funding from the European Research Council (ERC) under the European Unions Horizon 2020 research and innovation program (grant agreement No 803419). 

PB acknowledges financial support from the Poul Due Jensen Foundation: Financial support from the Poul Due Jensen Foundation (Grundfos Foundation) for this research is gratefully acknowledged. Grant number 2018-039.


\section*{Declaration of interest}
The authors report no conflicts of interest. 

\section*{Data availability}
The data that forms the basis of this study is available from the corresponding author upon reasonable request. 

\bibliography{aiptemplate_NSmachine}


\end{document}